\documentclass[11pt,twoside,a4paper]{article}

\usepackage[pdftex]{graphicx} %%%%%%%%%%%%%%%% paulo
\usepackage{amsmath,amssymb,amsthm}
\usepackage[ansinew]{inputenc}%%%%%%%%%%%%
\usepackage{a4wide}
\usepackage{xspace}%%%%%%%%%%%%%%%%%%%%%%

\theoremstyle{plain}
\newtheorem{thm}{Theorem}

\theoremstyle{definition}

\begin{document}
%\renewcommand{\baselinestretch}{1.31}%{1.31} Paulo
%\small\normalsize

\title{Two-dimensional body of maximum mean resistance\footnote{Accepted (April 16, 2009) for publication in the journal ``Applied Mathematics and Computation''.}}

\author{Paulo D. F. Gouveia\small ${\,}^a$ \normalsize\\
        \texttt{pgouveia@ipb.pt} \and
        Alexander Plakhov\small ${\,}^{b,c}$ \normalsize\\
        \texttt{plakhov@ua.pt} \and
        Delfim F. M. Torres\small ${\,}^{b}$ \normalsize\\
        \texttt{delfim@ua.pt}}

\date{\small ${}^a$Bragan\c{c}a Polytechnic Institute, 5301-854 Bragan\c{c}a, Portugal\\
${}^{b}$University of Aveiro, 3810-193 Aveiro, Portugal\\
${}^{c}$Aberystwyth University, SY23 3BZ Aberystwyth, UK}

\maketitle

\vspace{-1.0cm}

\begin{abstract}
A two-dimensional body, exhibiting a slight rotational movement, moves in a rarefied medium of particles which collide with it in a perfectly elastic way. In previously realized investigations by the first two authors, Plakhov \& Gouveia (2007, Nonlinearity, 20), shapes of nonconvex bodies were sought which would maximize the braking force of the medium on their movement. Giving continuity to this study, new investigations have been undertaken which culminate in an outcome which represents a large qualitative advance relative to that which was achieved earlier. This result, now presented, consists of a two-dimensional shape which confers on the body a resistance which is very close to its theoretical supremum value. But its interest does not lie solely in the maximization of Newtonian resistance; on regarding its characteristics, other areas of application are seen to begin to appear which are thought to be capable of having great utility. The optimal shape which has been encountered resulted from numerical studies, thus it is the object of additional study of an analytical nature, where it proves some important properties which explain in great part its effectiveness.
\end{abstract}

\medskip


\smallskip

%%%\oMSC 74F10, 65D15, 70E15, 49K30, 49Q10.

\smallskip

%%%%%%%%%%%%%%%%%%%%%%%%%%%%%%%%%%%%%%%%%%%%%%%%%%%%%%%%%%%%%%%%%%%%

\medskip

%%%%%%%%%%%%%%%%%%%%%%%%%%%%%%%%%%%%%%%%%%%%%%

\section{Introdution}
One area of investigation in contemporary mathematics is concerned with the search for shapes of bodies, within predefined classes, which permit the minimization or maximization of the resistance to which they are subjected when they move in rarefied media. The first problem of this nature goes back to the decade of the 1680s, a time when Isaac Newton studied a problem of minimum resistance for a specific class of convex bodies, which moved in media of infinitesimal particles, rarefied to such a degree that it was possible to discount any interaction between the particles, and in which the interaction of these with the bodies could be described as perfectly elastic collisions \cite{newton1686}. More recently we have witnessed important developments in this area with the broadening of study to new classes of bodies and to media with characteristics which are less restrictive: problems of resistance in non-symmetrical bodies~\cite{buttazzo95,buttazzo97,buttazzo93,robert06,robert01b}, in nonconvex bodies of single collisions~\cite{brock96,buttazzo93,comte01,robert01} and multiple collisions~\cite{Plakhov03b,plakhov03,Plakhov04}, bodies of developable surfaces~\cite{robert01}, considering collisions with friction~\cite{horstmann02} and in media with positive temperature~\cite{Plakhov05}. However most studies which have been published have given special attention to classes of convex bodies.

The convexity of a body is a sufficient condition for the resistance to be solely a function of singular collisions --- all the particles collide at once with the body. This attribute allows us to considerably reduce the complexity of the problems dealt with. Even the various studies of classes of nonconvex bodies which have emerged, especially in the last decade, are based almost always on conditions that guarantee a single impact per particle ---~\cite{brock96,buttazzo93,comte01,robert01}. Only very recently have there begun to emerge some studies supposing multiple reflections (see, e.g.~\cite{Plakhov03b,plakhov03,Plakhov04}).

In the class of convex bodies, the problem is normally reduced to the minimization of Newton's functional --- an analytical formula for the value of the resistance. But, in the context of nonconvex bodies, there is not any simple formula known for the calculation of the resistance. Even if it is extremely complex, in general, to deal analytically with problems of multiple collisions, for some specific problems of minimization the job has not been revealed to be particularly difficult, there even being some results already available~\cite{Plakhov03b,plakhov03}. If, on the other hand, we consider the problem of maximization, then in this case the solution becomes trivial --- for any dimension, it is enough that the front part of the body is orthogonal to the direction of the movement.

And what if the body exhibits, besides its translational movement, a slight rotational movement? When we think of this kind of problem, we have in mind, for example, artificial satellites, of relatively low orbits, which do not possess any control system which could stabilize their orientation, or other devices in similar conditions. In this situation we imagine that, over its path, the device rotates slowly around itself.

The problem of resistance minimization for rotating nonconvex two-dimensional bodies has already been studied in~\cite{Plakhov04,ARMA}: it was shown that the maximal reduction of resistance, as compared with the convex
case, is approximately $1.22\%$.
In its turn, the problem of maximization of the average resistance of bodies in rotation is far from being trivial, in contrast with that which occurs when we deal with purely translational movement.
This class of problems was, therefore, the object of study of the work carried out by the authors in \cite{Plakhov07:CM, Plakhov07}: nonconvex shapes of bodies were investigated which would maximize the resistance that they would have to confront if they moved in rarefied media, and, simultaneously, exhibited a slight rotational movement. With the numerical study which was executed, various geometrical shapes were found which conferred on the bodies rather interesting values of resistance: but it was in later investigations, performed in the follow-up of this work, that the authors managed to arrive at the best of the results --- a two-dimensional shape which confers on the body a resistance very near to its maximum theoretical limit. It is this latest result which now is presented here.

The presentation of the work is organized in the following way. In section~\ref{sec:defBidimens}, we begin by defining, for the two-dimensional case, the problem of maximization, which is the object of the present study. Then, in section~\ref{sec:estNumer}, we describe the numerical study which was realized in the tracking of the body of maximum resistance and we present the main original result of this study: a two-dimensional shape which maximizes Newtonian resistance. The two-dimensional shape is then the object of study in section~\ref{sec:caract}, where some important properties are shown which help to explain the value of resistance which it displays.
%%In section~\ref{sec:outrasAplic}, we put forth an exploratory study of other possible applications of our result.
In section~\ref{sec:concl}, we present the main conclusions of our study and include some notes on possible working directions to undertake in the future. Finally, in appendices~\ref{cha:condSuf3col} and~\ref{cha:min3col}, proofs of theorems~\ref{teor:3ref} and~\ref{teor:min3ref} are provided.

\section{Definition of the problem for the two-dimensional case}
\label{sec:defBidimens}
Consider a disc in slow and uniform rotation, moving in a direction parallel to its plane. We will designate the disc of radius $r$ by $C_r$ and its boundary by $\partial C_r$. We then remove small pieces of the disc along its perimeter, in an $\varepsilon$-neighborhood of $\partial C_r$, with $\varepsilon \in \mathbb{R}_+$ of value arbitrarily small when compared with the value of $r$. We are thus left with a new body $B$ defined by a subset of $C_r$ and characterized by a certain roughness along all its perimeter. The essential question which we put is the following: up to what point can the resistance of a body $B$ be augmented? More than getting to know the absolute value of this resistance, we are principally interested in learning what is the increase which can be obtained in relation to the smooth body (a perfectly circular contour, in this case), that is, learning the normalized value 
\begin{equation}
\label{eq:normalizacao}
R(B)=\frac{\text{Resistance}(B)}{\text{Resistance}(C_r)} \text{.}
\end{equation}
It is possible, from the beginning, to know some important reference values for the normalized resistance: $R(C_r)=1$ and the value of the resistance $R(B)$ will have to be found between $0.9878$ (\cite{Plakhov04,ARMA}) and $1.5$. The value $1.5$ will be hypothetically achieved if all the particles are reflected by the body with the velocity $\mathbf{v}^+$ (velocity with which the particles separate definitively from the body) opposite to the velocity of incidence $\mathbf{v}$ (velocity with which the particles strike the body for the first time), $\mathbf{v}^+=-\mathbf{v}$, the situation in which the maximum momentum is transmitted to the body. It is also possible for us to know the resistance value of some elementary bodies of the type $B$. This is the case, for example, of discs with the contour entirely formed by rectangular indentations which are arbitrarily small or with the shape of rectangular isosceles triangles. As was proved in \cite{Plakhov07}, these bodies are associated with resistances, respectively, of $R=1.25$ and $R=\sqrt{2}$.

Apart from being defined in the disc $C_r$, it is assumed that the body to be maximized is a connected set $B\in\mathbb{R}^2$, with piecewise smooth boundary $\partial B$.
Therefore, let us consider a billiard in $\mathbb{R}^2\setminus B$. An infinitesimal particle moves freely, until, upon colliding with the body $B$, it suffers various reflections (one at least) at regular points of its boundary $\partial B$, ending up by resuming free movement which separates it definitively from the body.
Denote by $\text{conv}B$ the convex hull of $B$.
The particle intercepts the $\partial(\text{conv}B)$ contour twice: when it enters into the set $\text{conv}B$ and in the moment that it leaves. $L=|\partial(\text{conv}B)|$ is considered the total length of the curve $\partial(\text{conv}B)$, and the velocity of the particle is in the first and second moments of interception represented by $\mathbf{v}$ and $\mathbf{v}^+$, and $x$ and $x^+$ the respective points where they occur. As well, the angles which the vectors $-\mathbf{v}$ and $\mathbf{v}^+$ make with the outer normal vector to the section of $\partial(\text{conv}B)$ between the points $x$ and $x^+$ are designated $\varphi$ and $\varphi^+$.
They will be positive if they are defined in the anti-clockwise direction from the normal vector, and negative in the opposite case. With these directions, both $\varphi$ as well as $\varphi^+$ take values in the interval $[-\pi/2,\pi/2]$.

\begin{figure}[!ht]
\begin{center}
\includegraphics[width=0.35\columnwidth]{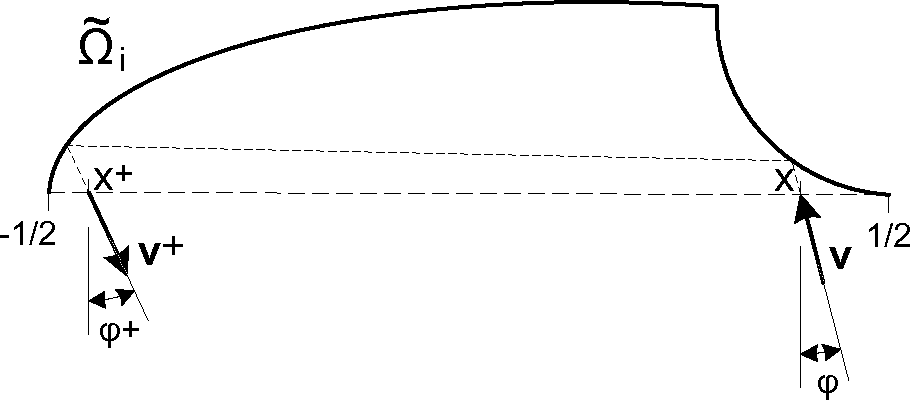}
\caption{Example of trajectory of a particle which interacts with a cavity $\tilde{\Omega}_i$.}
\label{fig:cavidadeOmega}
\end{center}
\end{figure}

Representing the cavities which characterize the contour of $B$ by subsets $\Omega_1,\Omega_2, \ldots$, which in their total make up the set $\text{conv}B \setminus B$, the normalized resistance of the body $B$ (equation~\eqref{eq:normalizacao}) takes the following form (cf. \cite{Plakhov07}): 
\begin{equation}
\label{eq:RB2}
R(B)
=\frac{|\partial(\text{conv}B)|}{|\partial C_r|} \left(\frac{L_0}{L}+\sum_{i \ne 0}{\frac{L_i}{L}R(\tilde{\Omega}_i)}\right)
\text{,}
\end{equation}
being $L_0=|\partial(\text{conv}B)\cap \partial B|$ the length of the convex part of the contour $\partial B$, $L_i=|\partial(\text{conv}B)\cap {\Omega}_i|$, with $i=1,2,\ldots$, the size of the opening of the cavity ${\Omega}_i$, and $R(\tilde{\Omega}_i)$ the resistance of the normalized cavity $\tilde{\Omega}_i$, in relation to a smooth segment of unitary size, with
\begin{equation}
\label{eq:R}
R(\tilde{\Omega}_i)=\frac{3}{8} \int_{-1/2}^{1/2}\int_{-\pi/2}^{\pi/2} \left(
1+\cos\left( \varphi^+(x,\varphi) -\varphi \right) 
\right) \cos \varphi\, \mathrm{d}\varphi\,\mathrm{d} x
\text{.}
\end{equation}
The function $\varphi^+$ should be seen as the angle of departure of a particle which interacts with a cavity $\tilde{\Omega}_i$ that has opening of unit size and is similar to $\Omega_i$, with the similarity factor $1/L_i$ --- see illustration of figure~\ref{fig:cavidadeOmega}.

%%%%%%%%% figura %%%%%%%%%%%%%%%%%%%%%%%%%%%%%%%%%%%%%%%%%%%%%%%%%%%%%%%%%%%%%%
%%%%%%%%% figura %%%%%%%%%%%%%%%%%%%%%%%%%%%%%%%%%%%%%%%%%%%%%%%%%%%%%%%%%%%%%%

From equation~\eqref{eq:RB2}, we understand that the resistance of $B$ can be seen as a weighted mean ($\sum_i L_i/L=1$) of the resistances of the individual cavities which characterize all its boundary (including resistance of the convex part of the boundary), multiplied by a factor which relates the perimeters of the bodies $\text{conv}B$ and $C_r$. Thus, maximizing the resistance of the $B$ body amounts to maximizing the perimeter of $\text{conv}B$ ($|\partial(\text{conv}B)|\le |\partial C_r|$) and the individual resistances of the cavities $\Omega_i$. 

Having found the optimal shape $\Omega^*$, which maximizes the functional \eqref{eq:R}, the body of maximum resistance $B$ will be that whose boundary is formed only by the concatenation of small cavities with this shape. We can therefore restrict our problem to the sub-class of bodies $B$ which have their boundary integrally covered with equal cavities, and in doing so admit, without any loss of generality, that each cavity $\Omega_i$ occupies the place of a circle arc of size $\varepsilon\ll r$. As with $L_i=2r\sin(\varepsilon/2r)$, the ratio between the perimeters takes the value
\begin{equation}
\label{eq:RxPerimetros}
\frac{|\partial(\text{conv}B)|}{|\partial C_r|}
=\frac{ \sin(\varepsilon/2r)}{\varepsilon/2r}
\approx 1- \frac{(\varepsilon/r)^2}{24}\text{,}
\end{equation}
or that is, given a body $B$ of a boundary formed by cavities similar to $\Omega$, from~\eqref{eq:RB2} and~\eqref{eq:RxPerimetros}, we conclude that the total resistance of the body will be equal to the resistance of the individual cavity $\Omega$, less a small fraction of this value, which can be neglected when $\varepsilon\ll r$,
\begin{equation}
\label{eq:Raprox}
R(B)\approx R(\Omega)-\frac{(\varepsilon/r)^2}{24}R(\Omega)\text{.}
\end{equation}

Thus, our research has as its objective the finding of cavity shapes $\Omega$ which maximize the value of the functional~\eqref{eq:R}, whose limit we know to be found in the interval
\begin{equation}
\label{eq:RminMax}
1\le \text{sup}_\Omega R(\Omega) \le 1.5
\text{,}
\end{equation}
as is easily proven using~\eqref{eq:R}:
if $\Omega$ is a smooth segment, $\varphi^+(x,\varphi)=-\varphi$ and
$R(\Omega)=\frac{3}{8} \int_{-1/2}^{1/2}\int_{-\pi/2}^{\pi/2} \left(
1+\cos\left(  2\varphi \right) 
\right) \cos \varphi\, \mathrm{d}\varphi\,\mathrm{d} x=1
$;
in the conditions of maximum resistance $\varphi^+(x,\varphi)$ $=\varphi$,
thus $
R(\Omega)\le\frac{3}{8} \int_{-1/2}^{1/2}\int_{-\pi/2}^{\pi/2} 2 \cos \varphi\, \mathrm{d}\varphi\,\mathrm{d} x
=1.5$.

\section{Numerical study of the problem}
\label{sec:estNumer}
In the class of problems which we are studying, only for some shapes of $\Omega$ which are very elementary is it possible to derive an analytical formula of their resistance~\eqref{eq:R}, as we saw in the rectangular and triangular shapes previously referred to. For somewhat more elaborate shapes, the analytical calculation becomes rapidly too complex, if not impossible, given the great difficulty in knowing the function $\varphi^+: [-1/2,1/2]\times[-\pi/2,\pi/2]\rightarrow[-\pi/2,\pi/2]$, which as we know, is intimately related to the format of the cavity $\Omega$. Therefore, recourse to numerical computation emerges as the natural and inevitable approach in order to be able to investigate this class of problems.

There have been developed various computational models which simulate the dynamics of billiard in the cavity. The algorithms of construction of these models, as well as the those responsible for the numerical calculation of the associated resistance, were implemented using the programming language C, given the computational effort involved (language C was created in 1972 by Dennis Ritchie; for its study we suggest, among the extensive documentation available, that which is the reference book of its language, written by Brian Kernighan and Dennis Ritchie himself, \cite{LangC}). The efficiency of the object code, generated by the compilers of C, allowed the numerical approximation of~\eqref{eq:R} to be made with a sufficiently elevated number of subdivisions of the intervals of integration --- between some hundreds and various thousands (up to $5000$). The results were, because of this, obtained with a precision which reached in some cases $10^{-6}$. This precision was controlled by observation of the difference between successive approximations of the resistance $R$ which were obtained with the augmentation of the number of subdivisions.

For the maximization for the resistance of the idealized models, there were used the global algorithms of optimization of the \textit{toolbox} ``\emph{Genetic Algorithm and Direct Search}''  (version 2.0.1 (R2006a), documented in~\cite{toolbox}), a collection of functions which extends the optimization capacities of the MATLAB numerical computation system. % \textregistered{}
The option for Genetic and Direct search methods is essentially owed to the fact that these do not require any information about the gradient of the objective function nor about derivatives of a higher order --- as the analytical form of the resistance function is in general unknown (given that it depends on $\varphi^+(x,\varphi)$), this type of information, if it were necessary, would have to be obtained by numerical approximation, something which would greatly impede the optimization process. The MATLAB computation system (version 7.2 (R2006a)) was also chosen because it had functionalities which allowed it to be used for the objective function the subroutine compiled in C of resistance calculation, as well as the $\varphi^+(x,\varphi)$ function invoked in itself.
%%%%%%%%%%%%%%%%%%%%%%%%%%%%%%%%%%%%%%%%%%%%%%

%%%%%%%%%%%%%%%%%%%%%%%%%%%%%%%%%%%%%%%%%%%%%%
\subsection{``Double Parabola'': a two-dimensional shape which maximizes resistance}
\label{sec:DuplaParab}
In the numerical study which the authors carried out in \cite{Plakhov07:CM, Plakhov07}, shapes of $\Omega_f$ defined by continuous and piecewise differentiable $f:[-1/2,1/2]\rightarrow \mathbb{R}^+$ functions were sought for:
\begin{equation}
\label{eq:Omegaf}
\Omega_f=\left\{(x,y):\,-1/2\le x \le 1/2,\; 0\le y \le f(x)\right\}\text{,}
\end{equation}
with the interval $[-1/2, 1/2] \times \{ 0 \}$ being the opening.

The search for the maximum resistance was begun in the class of continuous functions $f$ with derivative $f'$ piecewise constant, broadening later to the study of classes of functions with the second derivative $f''$ piecewise constant. In the first of the cases the contour of $\Omega_f$ is a polygonal line, and in the second, a curve composed of parabolic arcs. Not having been able with these shapes to exceed the value of resistance $R=1.44772$, we decided, in this new study, to extend the search to shapes different from those considered in~\eqref{eq:Omegaf}. We studied shapes $\Omega^g$ defined by functions $x$ of $y$ of the following form:
\begin{equation}
\label{eq:Omegag}
\Omega^g=\left\{(x,y):\,0\le y \le h,\; -g(y)\le x \le g(y)\right\}\text{,}
\end{equation}
where $h>0$ and $g:[0,h] \rightarrow \mathbb{R}_0^+$ is a continuous function with $g(0)=1/2$ and $g(h)=0$.

The new problem of maximum resistance studied by us can therefore be formulated in the following way:
\begin{quotation}{\sl
To find $\sup_g R(\Omega^g)$ in the continuous and piecewise differentiable functions
$g:[0,h] \rightarrow \mathbb{R}_0^+$,
such as $g(0)=1/2$ and $g(h)=0$, with $h>0$.
}\end{quotation}

Similarly to the study which was carried out for the sets $\Omega_f$, in the search for shapes $\Omega^g$, the functions $g$ were considered piecewise linear and piecewise quadratic. 
If in the classes of linear functions it was not possible to achieve a gain in resistance relative to the results obtained for the sets $\Omega_f$, in the quadratic functions the results exceeded the highest expectations: there was found a shape of cavity $\Omega^g$ which presented the resistance $R=1.4965$, a value very close to its theoretical limit of $1.5$.
%This is surely a very interesting result.
There were also carried out some tests with polynomial functions of higher order or described by specific conical sections, but, not having verified any additional gain in the maximization of resistance, it was decided not to report the respective results. There therefore follows the description of the best result which was obtained, encountered in the class of quadratic functions $x=\pm g(y)$.

\begin{figure}[!ht]
\begin{center}
\begin{tabular}{c c}
\includegraphics*[height=0.3\columnwidth]{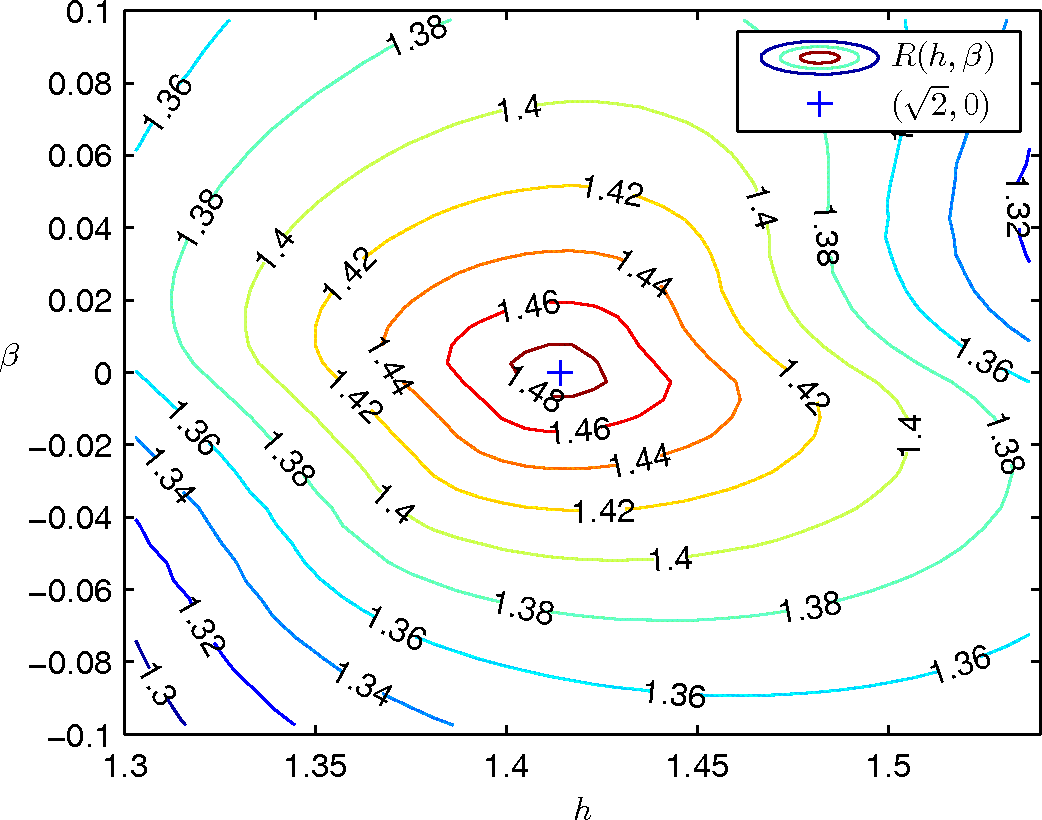}
&
\includegraphics*[height=0.3\columnwidth]{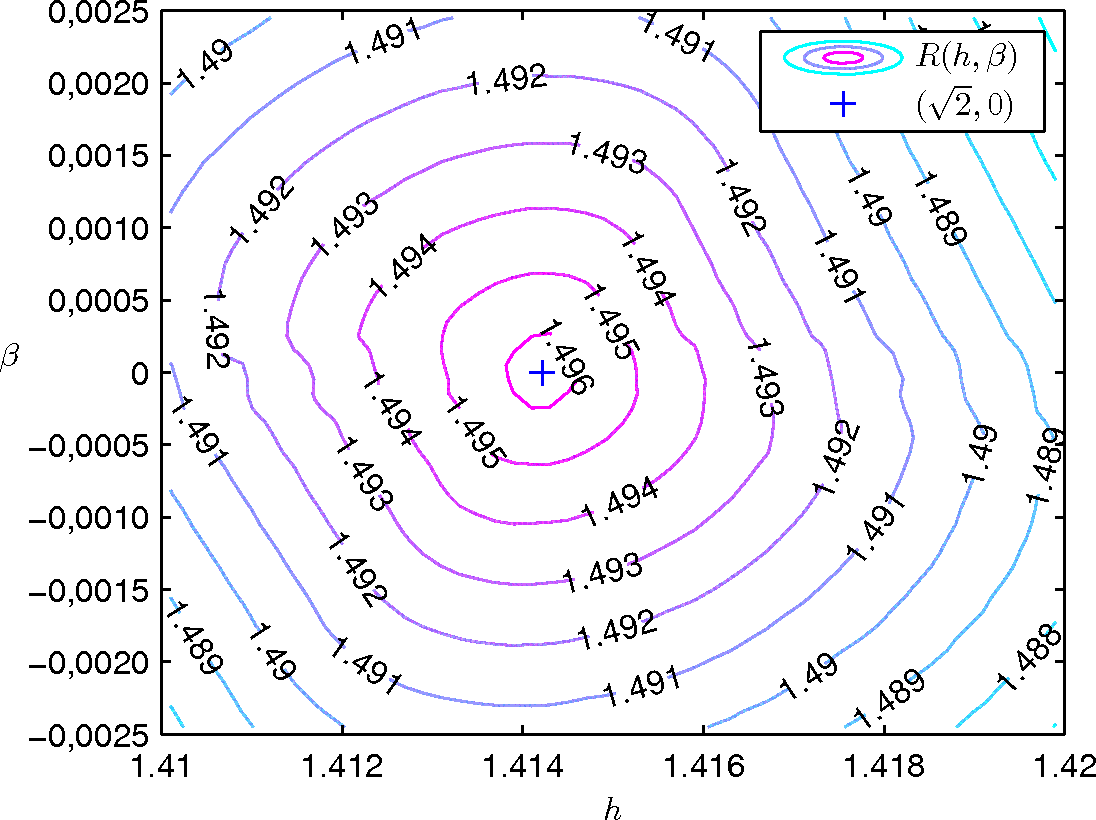} \\
(a)&(b)\\
\end{tabular}
\caption{Level curves of the $R(h,\beta)$ function.}
\label{fig:curvNivel}
\end{center}
\end{figure}

\begin{figure}[!ht]
\begin{center}
\includegraphics*[width=0.4\columnwidth]{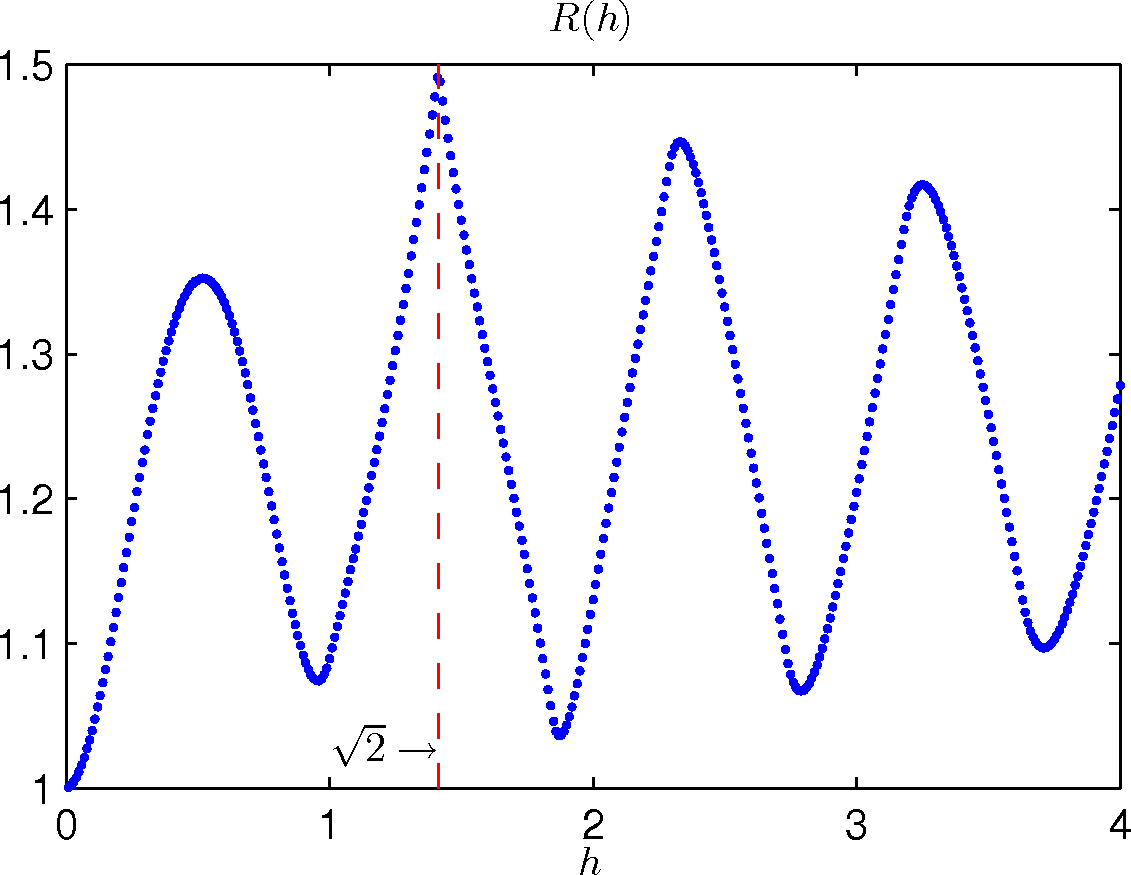}
\caption{Resistance graphic $R(h)$ for $\beta=0$.}
\label{fig:Res_h}
\end{center}
\end{figure}

%\subsubsection*{Funções $x=\pm g(y)$ quadráticas}
The value of resistance of the sets $\Omega^g$ were studied, just as defined in~\eqref{eq:Omegag}, in the class of quadratic functions
$$
g_{h,\beta}(y) = 
 \alpha y^2 + \beta y +1/2, \text{ for } 0 \le y \le h\,,
$$
where $h>0$ and $\alpha= \frac{-\beta h -1/2}{h^2}$ (given that $g_{h,\beta}(h) =0$). In the optimization of the curve, the two parameters of the configuration were made to vary: $h$, the height of the $\partial\Omega^g$ curve, and $\beta$, in its slope at the origin ($g'(0)$). In this class of functions the algorithms of optimization converge rapidly towards a very interesting result:
the maximum resistance was reached with $h=1.4142$ and $\beta=0.0000$, and assumed the value $R=1.4965$, that is, a value $49.65\%$ above the resistance of the rectilinear segment.
%It is really a very important result:
This result seems to us really interesting:
\begin{enumerate}
\item it represents a considerable gain in the value of the resistance, relative to the best result obtained earlier (in \cite{Plakhov07:CM, Plakhov07}), 
which was situated $44.77\%$ above the reference value;
\item The corresponding set $\Omega^g$ has a much more simple shape than that of set $\Omega_f$ associated with the best earlier result, since it is formed by two arcs of symmetrical parabolas, while the earlier one was made up of fourteen of these arcs;
\item this new resistance value is very near to its maximal theoretical limit, which, as is known, is found $50\%$ above the value of reference;
\item The optimal parameters appear to assume value which give to the set $\Omega^g$ a configuration with very special characteristics, as in what follows will be understood.
\end{enumerate}
Note that the optimal parameters appear to approximate the values $h=\sqrt{2}=1.41421\ldots$ and $\beta=0$. The following question can therefore be put:
\begin{quotation}{\sl
Are these not the exact values of the optimal parameters?
}\end{quotation}
The graphical representation of the function $R(h,\beta)$  through the level curves, figure~\ref{fig:curvNivel}, are effectively in concordance with this possibility
 --- note that the level curves appear perfectly centered on the $(\sqrt{2},0)$ coordinates; marked on the figure by ``$+$'' .
%%%%%%%%% figura %%%%%%%%%%%%%%%%%%%%%%%%%%%%%%%%%%%%%%%%%%%%%%%%%%%%%%%%%%%%%%
%%%%%%%%% figura %%%%%%%%%%%%%%%%%%%%%%%%%%%%%%%%%%%%%%%%%%%%%%%%%%%%%%%%%%%%%%
Note also, in figure~\ref{fig:Res_h}, the resistance graph $R(h)$ for $\beta=0$, where it can equally be perceived that there is a surprising elevation of resistance when $h\rightarrow\sqrt{2}$. Thus the resistance of the $\Omega^{g_{h\beta}}$ cavity was numerically calculated with the exact values $h=\sqrt{2}$ and $\beta=0$, the result having confirmed the value $1.49650$.
%%%%%%%%% figura %%%%%%%%%%%%%%%%%%%%%%%%%%%%%%%%%%%%%%%%%%%%%%%%%%%%%%%%%%%%%%
%%%%%%%%% figura %%%%%%%%%%%%%%%%%%%%%%%%%%%%%%%%%%%%%%%%%%%%%%%%%%%%%%%%%%%%%%

\begin{figure}[!ht]
\begin{center}
\begin{tabular}{c c c}
\includegraphics*[height=0.25\columnwidth]{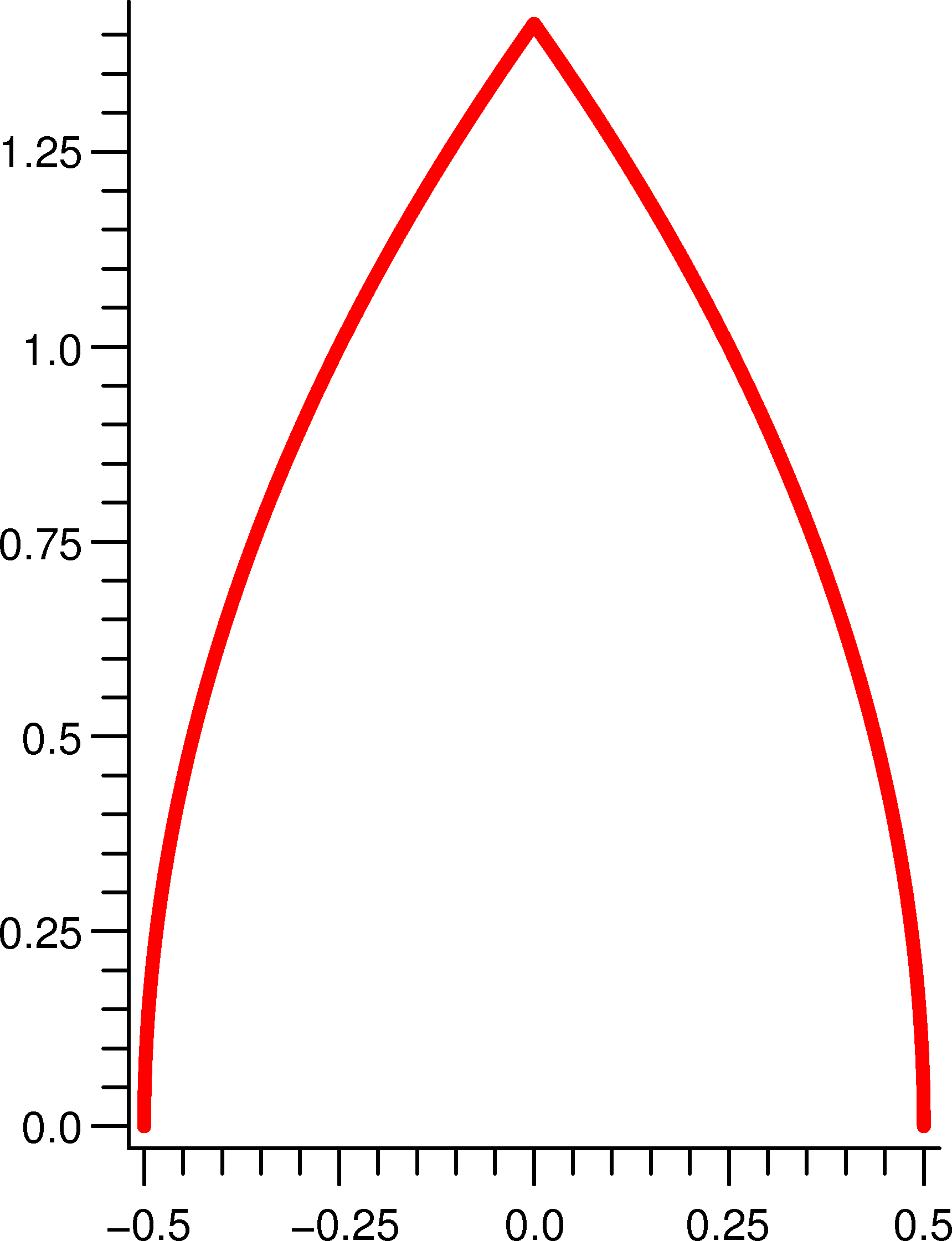}
&
&
\includegraphics*[height=0.25\columnwidth]{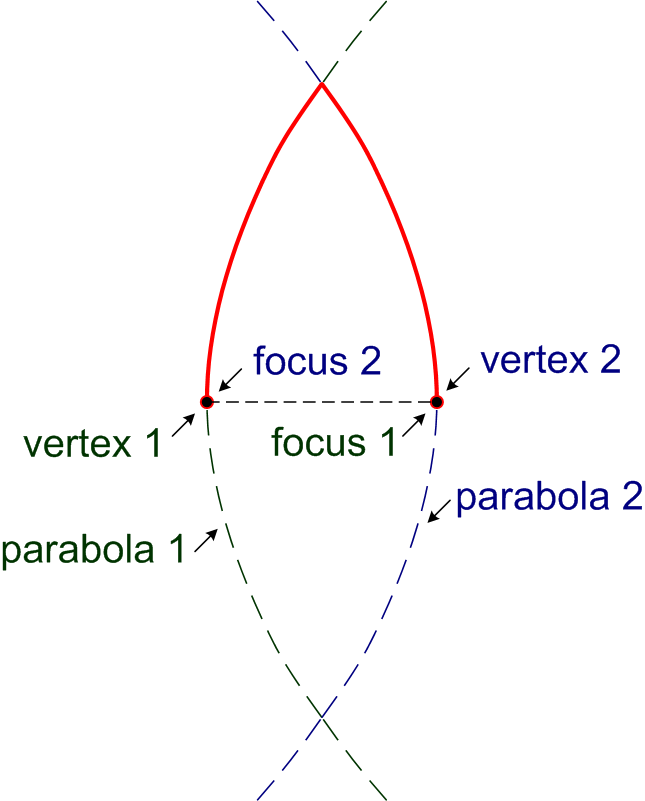} \\
(a)&&(b)\\
\end{tabular}
\caption{(almost) Optimal 2D shape --- the \emph{Double Parabola}.}
\label{fig:parabOpt}
\end{center}
\end{figure}

There is yet one more reason which suggests also an affirmative response to the formulated question. The shape of the set $\Omega^{g_{h,\beta}}$ with $h=\sqrt{2}$ and $\beta=0$ is a particular case with which is associated special characteristics which could justify the elevated value of resistance presented. The two sections of the shape are similar arcs of two parabolas with the common horizontal axis and concavities turned one towards the other --- see figure~\ref{fig:parabOpt}. But the particularity of the configuration resides in the fact that the axis of the parabolas coincides with the line of entry of the cavity (axis of $x$), and that the focus of each one coincides with the vertex of the other.
%Note that, the equation of the parabola on the right side being given as $x=g_{\sqrt{2},0}(y)$, $y^2=-4(x-1/2)$ is obtained, confirming that the vertex is in $(1/2,0)$ and that the focus is found one unit from this. In this way, for reasons of symmetry, the foci and the vertices of the two parabolas are found at the extremes of the opening of the cavity, $(-1/2,0)$ and $(1/2,0)$, just as is illustrated in the scheme of figure~\ref{fig:parabOpt}b.
%%%%%%%%% figura %%%%%%%%%%%%%%%%%%%%%%%%%%%%%%%%%%%%%%%%%%%%%%%%%%%%%%%%%%%%%%
%%%%%%%%% figura %%%%%%%%%%%%%%%%%%%%%%%%%%%%%%%%%%%%%%%%%%%%%%%%%%%%%%%%%%%%%%

%The elevated value of resistance of this shape of cavity is also expressed, although implicitly, in the graph of the integrand function of functional~\eqref{eq:R}, presented in figure~\ref{fig:funInteg}. It displays a surface with a rather smooth shape over almost all its domain and, most importantly, nearly does not vary with $x$ and appears to describe the form of function $2\cos\varphi$ in the direction of the axis of the $\varphi$. These attributes, when verified in their fullness, are associated with the greatest admissible value of resistance, $R=1.5$.\footnote{If the integrand function takes the form $G(x,\varphi)=2\cos\varphi$, from~\eqref{eq:R} is obtained $R=1.5$.}
%%%%%%%%% figura %%%%%%%%%%%%%%%%%%%%%%%%%%%%%%%%%%%%%%%%%%%%%%%%%%%%%%%%%%%%%%
%\begin{figure}[!ht]
%\begin{center}
%\includegraphics*[width=0.4\columnwidth]{fig5}
%\caption{Graph of the integrand function $G(x,\varphi)=\left(
%1+\cos\left( \varphi^+(x,\varphi) -\varphi \right) 
%\right) \cos \varphi$.}
%\label{fig:funInteg}
%\end{center}
%\end{figure}
%%%%%%%%% figura %%%%%%%%%%%%%%%%%%%%%%%%%%%%%%%%%%%%%%%%%%%%%%%%%%%%%%%%%%%%%%

This shape of cavity appears to effectively deal with a very particular case. In contrast with what happened with all the other shapes which were studied, the integrand function of functional~\eqref{eq:R} displays a rather smooth shape, presenting only a few small irregularities for $\varphi$ angles of little amplitude. Noting this characteristic, and taking into account that the integrand function almost does not depend on $x$, the resistance was calculated, for this shape in particular, using the rule of Simpson $1/3$ in the integration in order to $\varphi$. The double integration in the equation~\eqref{eq:R} was thus numerically approximated by the following expression:
\begin{equation}
\label{eq:RnsSimpson}
R=\frac{1}{2} \Delta x \Delta \varphi \sum_{i=N_x/2+1}^{N_x}\sum_{k=1}^{N_\varphi-1} w_k \left(
1+\cos\left( \varphi^+(x_i,\varphi_k) -\varphi_k \right) 
\right) \cos \varphi_k\text{,}
\end{equation}
with $w_k=2$ for $k$ odd and $w_k=1$ for $k$ even, $x_i=- 1/2+(i-1/2) \Delta x$, $\Delta x=1/N_x$, $\varphi_k=-\pi/2+ k \Delta \varphi$ and $\Delta \varphi=\pi/N_\varphi$. $N_x$ and $N_\varphi$ are the number of sub-intervals to consider in the integration of the variables $x$ and $\varphi$ (both even numbers), respectively, and $\Delta x$ and $\Delta \varphi$ the increments for the correspondent discreet variables. Given that the shape $\Omega^{g_{\sqrt{2},0}}$ presents horizontal symmetry, the first summation of the expression considers only the second half of the interval of integration of the variable $x$.

In order to be easily referred to, this shape of cavity (figure~\ref{fig:parabOpt}a) will be, from here on, named simply ``\emph{Double Parabola}''. Thus, in the context of this paper, the term ``Double Parabola''  should be always understood as the name of the cavity whose shape is described by two parabolas which, apart from being geometrically equal, find themselves ``nested'' in the particular position to which we have referred.

Since the resistance of the Double Parabola assumes a value which is very close to its theoretical limit, in a final attempt to achieve this limit, it was resolved to extend the study even further to other classes of functions $g(y)$ which admit the Double Parabola as a particular case or which allow proximate configurations of this nearly optimal shape. In all these cases the best results were invariably obtained when the shape of the curves approximated the shape of the Double Parabola, without ever having overtaken the value $R=1.4965$.
It was begun by considering functions $g(y)$ piecewise quadratic, including curves \textit{splines}, without achieving interesting results; only for functions $g(y)$ of $2$ or $3$ segments was it possible to approximate the resistance and the shape of the Double Parabola.
Cubic and bi-quadratic functions $g(y)$ were also considered\footnote{In the bi-quadratic curves, the point of interception of the trajectory of the particle with the boundary of the cavity is calculated by resolving an equation of the 4th degree. The roots of this equation were obtained numerically using the method described in \cite{Hook90}. The equations of inferior order were always resolved by utilising the known analytical formulas.}, but in both cases the process of optimization brought them proximate to the curves of quadratic order, with the coefficients of greater order taking values which were almost zero.
The problem was studied in the class of conical sections, considering, for lateral facets of the cavity, two symmetrical arcs either of an ellipse or of a hyperbole. Also in these cases the arcs assumed a shape very close to the arcs of the parabolas.

The Double Parabola being the best shape encountered, and dealing with a nearly optimal shape, in the section which follows it is the object of deeper study, of an essentially analytical nature, where the reasons for its good performance are sought.

%%%%%%%%%%%%%%%%%%%%%%%%%%%%%%%%%%%%%%%%%%%%%%%%%%%%%%%%%%%%%%%%%%%%%%%%%%%%%%%%%%%
\section{Characterization of the reflections in the shape ``Double Parabola''}
\label{sec:caract}
\begin{figure}[!hb]
\begin{center}
\begin{tabular}{c c c}
\includegraphics*[width=0.15\columnwidth]{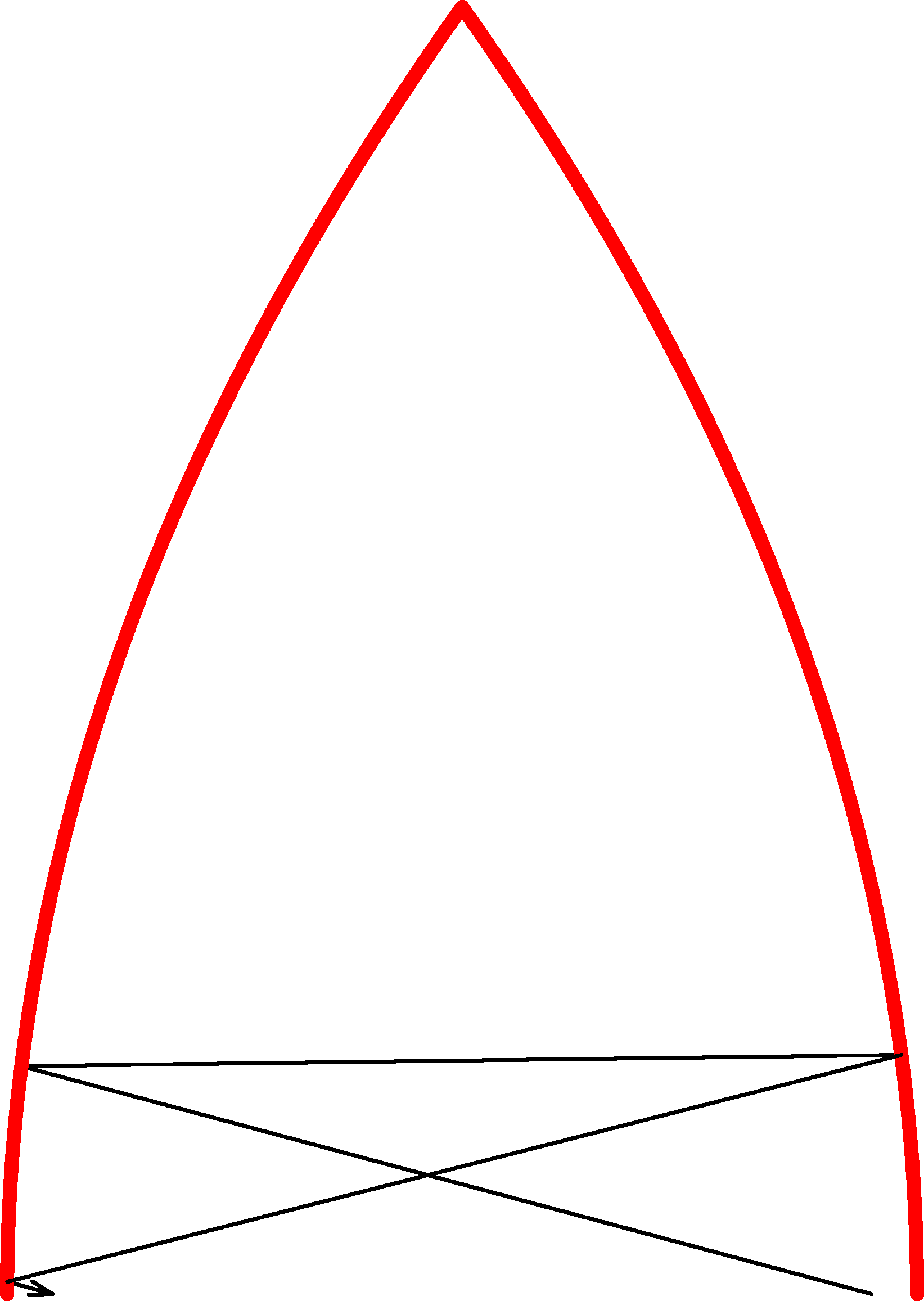} &
\includegraphics*[width=0.15\columnwidth]{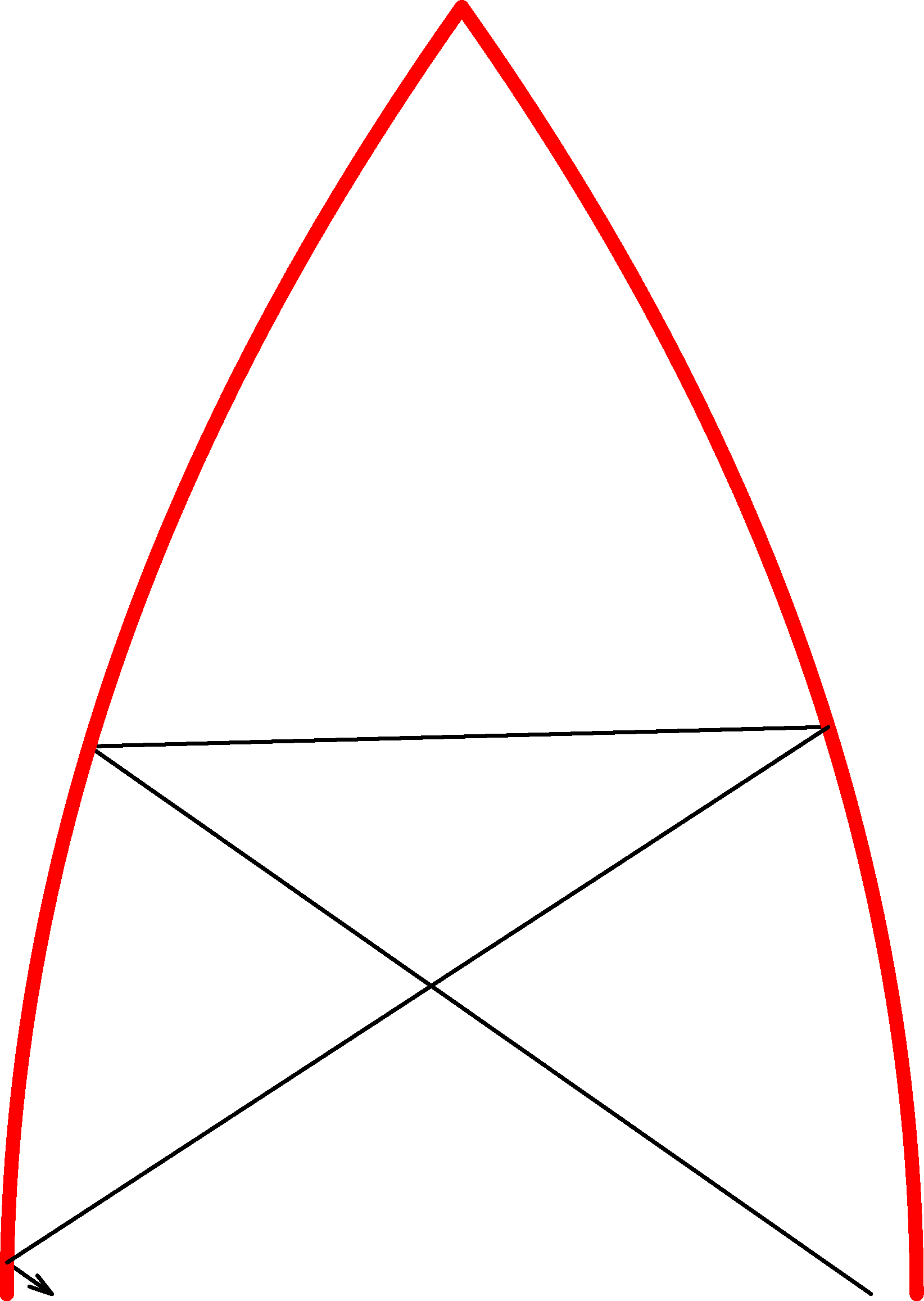} &
\includegraphics*[width=0.15\columnwidth]{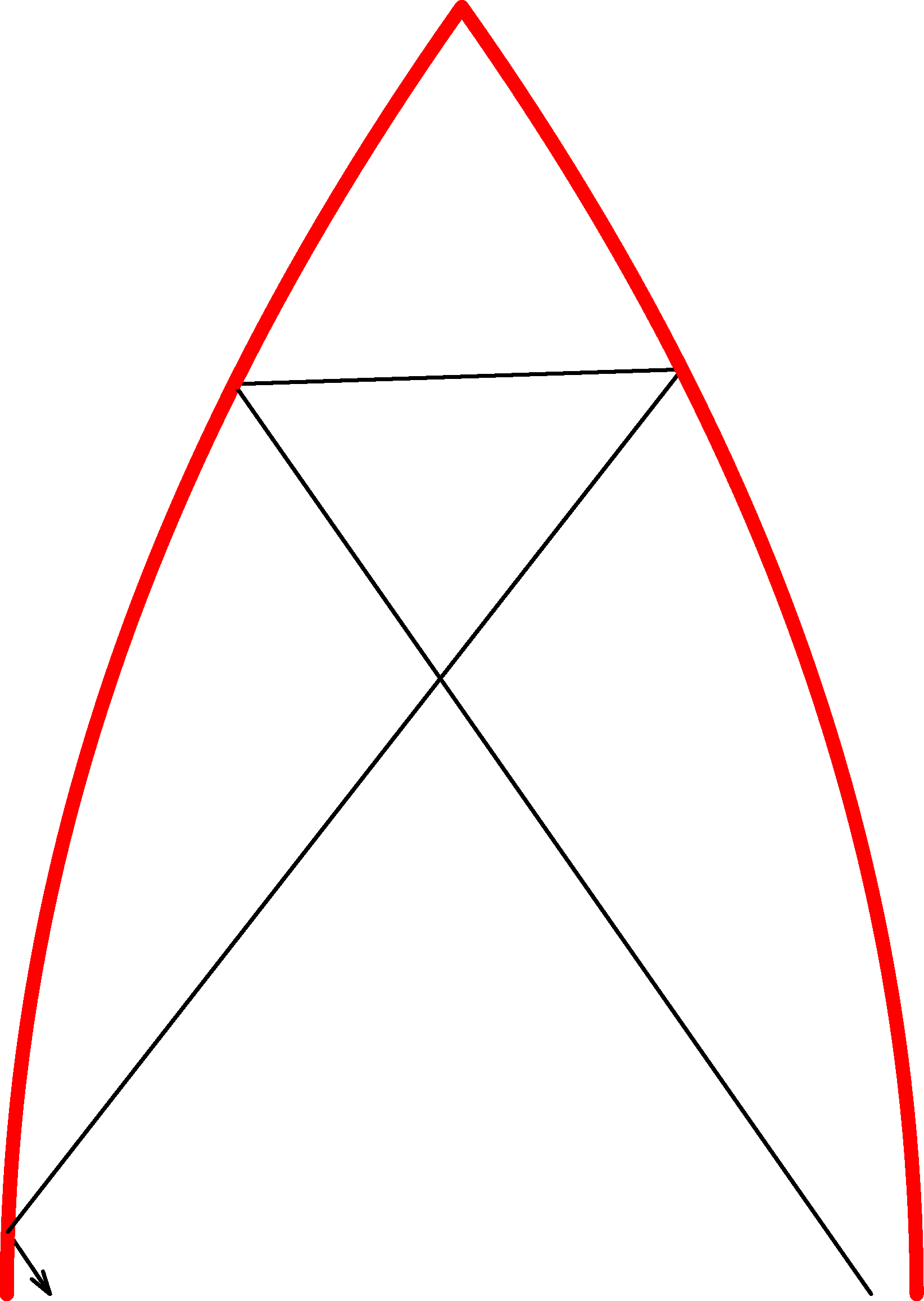} \\
(a) $x=0.45$, $\varphi=75^\circ$.&(b) $x=0.45$, $\varphi=55^\circ$.&(c) $x=0.45$, $\varphi=35^\circ$.\\
\includegraphics*[width=0.15\columnwidth]{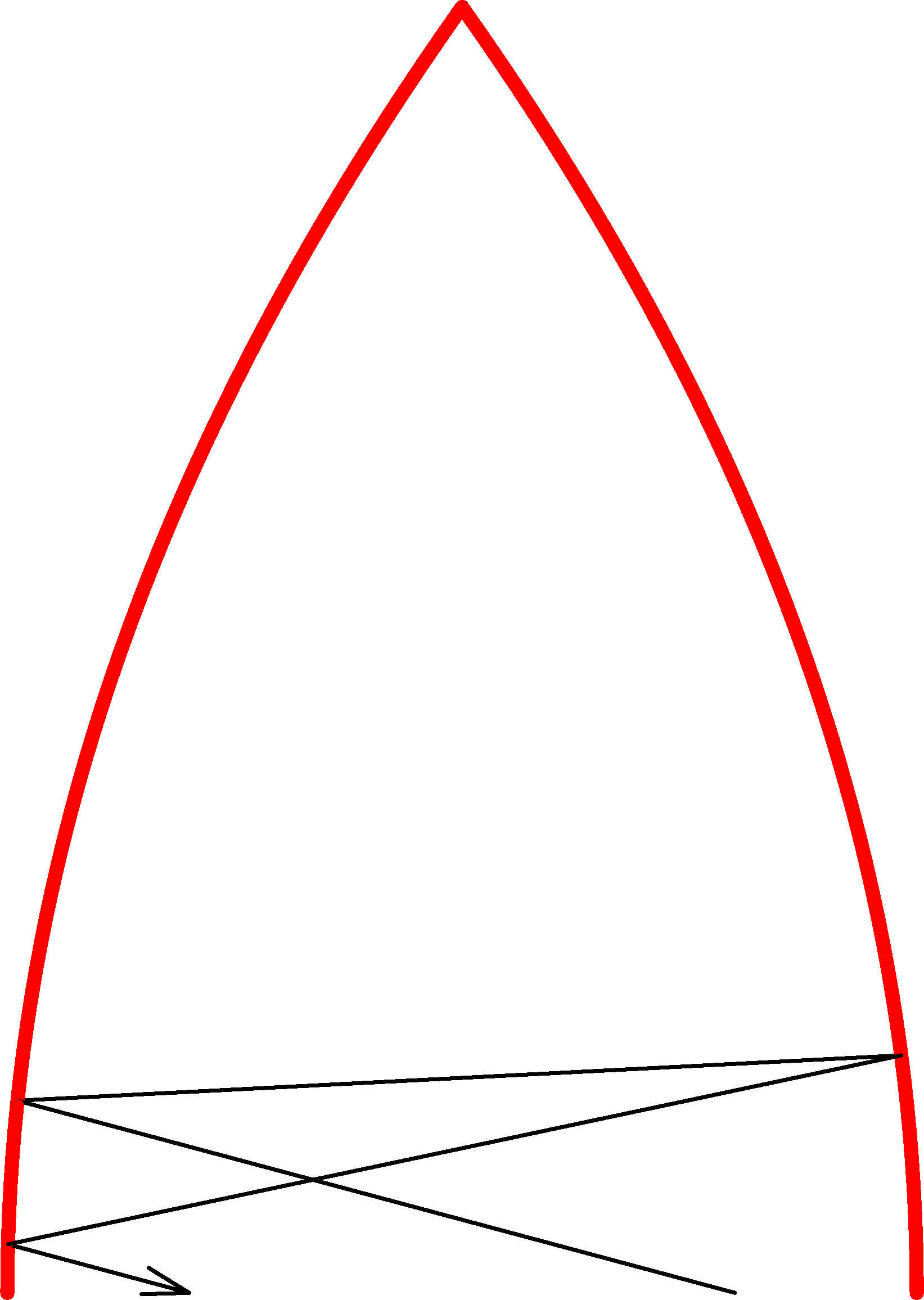} &
\includegraphics*[width=0.15\columnwidth]{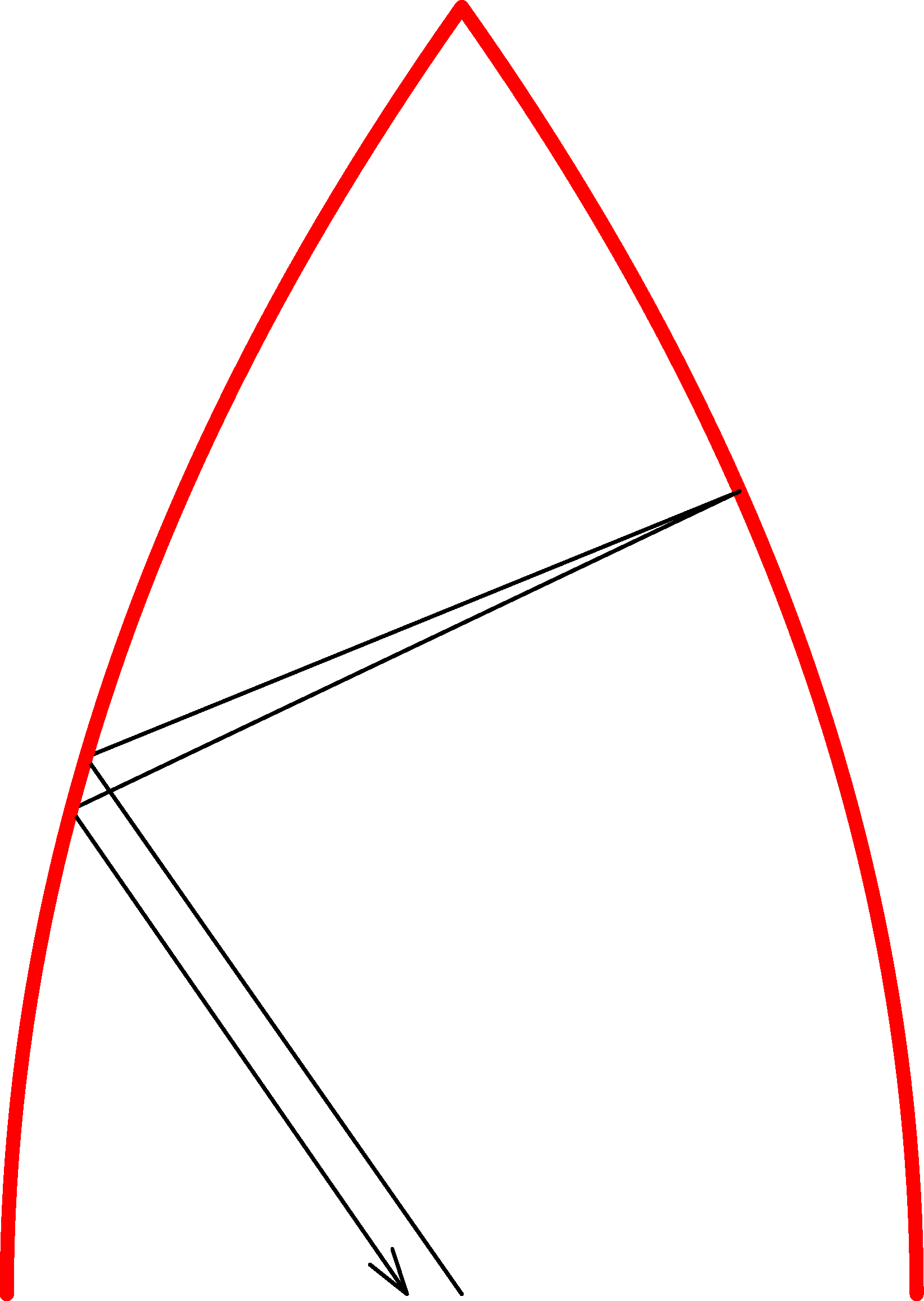} &
\includegraphics*[width=0.15\columnwidth]{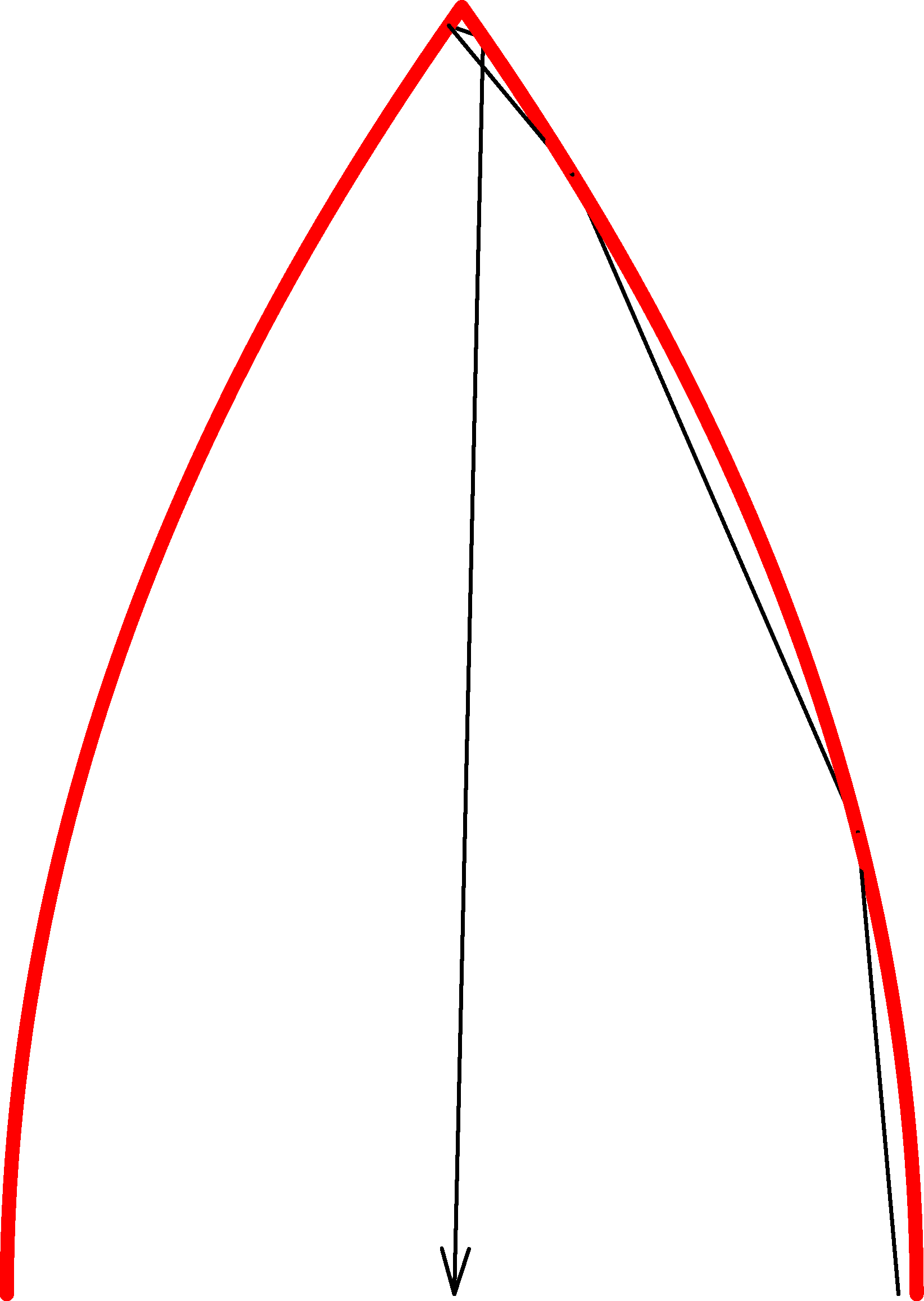} \\
(d) $x=0.3$, $\varphi=75^\circ$.&(e) $x=0.0$, $\varphi=35^\circ$.&(f) $x=0.48$, $\varphi=5^\circ$. \\
\end{tabular}
\caption{Example of trajectories obtained with the computational model.}
\label{fig:trajectorias}
\end{center}
\end{figure}

Each one of the illustrations of figure~\ref{fig:trajectorias} shows, for the ``Double Parabola'', a concrete trajectory, obtained with our computational model.
%%%%%%%%% Figura %%%%%%%%%%%%%%%%%%%%%%%%%%%%%%%%%%%%%%%%%%%%%%%%%%%%%%%%%%%%%%
%%%%%%%%% Figura %%%%%%%%%%%%%%%%%%%%%%%%%%%%%%%%%%%%%%%%%%%%%%%%%%%%%%%%%%%%%%
It is comforting to verify that, with the exception of one trajectory, in all the others the particle emerges from the cavity with a velocity which is nearly opposite to that which was its entry velocity. This is the ``symptom''  which unequivocally characterizes a cavity of optimal performance. Even in the case of the trajectory of the illustration (f), the direction of the exit velocity appears not to vary greatly from that of entry.

If we analyze the five first illustrations, we may verify that there exists something in common in the behavior of the particle: in describing the trajectory, the particle is always subject to three reflections. This appears to be a determinant characteristic for the approximation of the two angles of entry and exit. If, for example, we imagine three trajectories with proximate configurations, respectively, the trajectories (a), (b) and (c), but with the difference of not possessing the third reflection, the result would be completely different, as easily can be seen in the illustrations. Although this conviction is by nature essentially empirical, the results of the study which follow are heading in the direction of confirming that one very significant part of the ``benign''  trajectories --- those in which the vectors velocity of entry and of exit are nearly parallel; we call them so because they represent positive contributions to the maximization of resistance --- suppose exactly three reflections.

\begin{figure}[!hb]
\begin{center}
\includegraphics*[width=0.4\columnwidth]{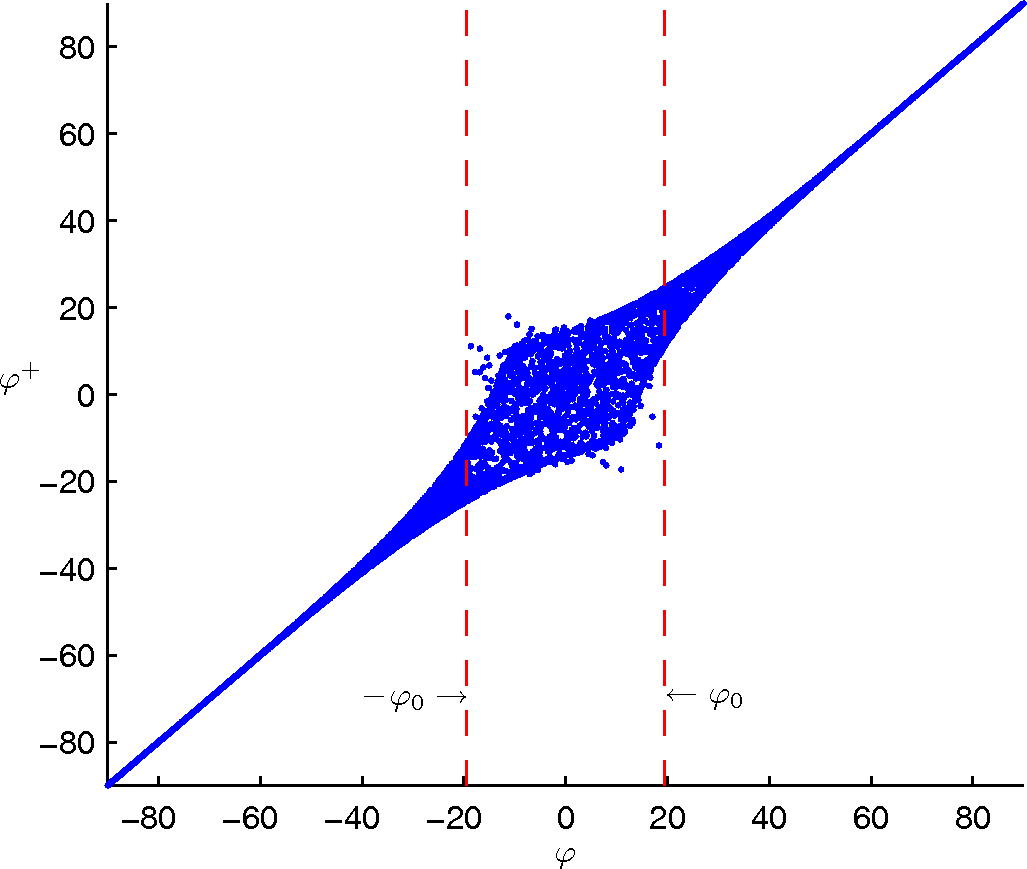}
\caption{Distribution of the $(\varphi,\varphi^+)$ pairs on the Cartesian plane.}
\label{fig:Dist_Phi_PhiPlus}
\end{center}
\end{figure}

We now will try to interpret another type of results obtained with our computational model, commencing with the graphical representation of the distribution of the pairs $(\varphi,\varphi^+)$ on the Cartesian plane --- see figure~\ref{fig:Dist_Phi_PhiPlus}. This graph was produced with $10.000$ pairs of values $(x,\varphi)$, generated by a random process of uniform distribution.

%%%%%%%%% Figura %%%%%%%%%%%%%%%%%%%%%%%%%%%%%%%%%%%%%%%%%%%%%%%%%%%%%%%%%%%%%%
%%%%%%%%% Figura %%%%%%%%%%%%%%%%%%%%%%%%%%%%%%%%%%%%%%%%%%%%%%%%%%%%%%%%%%%%%%
The points concentrate themselves on the proximities of the diagonal $\varphi=\varphi^+$, which revealing of good behavior on the part of the cavity. In addition, with these results it is shown that the response of the cavity deteriorates as $\varphi$ approaches zero. Therefore, it begins to be understood that the ``benign''  trajectories have their origin essentially in entry angles of elevated amplitude.

If we consider figure~\ref{fig:Dist_Phi_PhiPlus}, there appears to exist an additional perturbation in the behavior of the cavity when the amplitude of the entry angle is inferior to about $20^\circ$, which means that some $(\varphi,\varphi^+)$ pairs become, in relation to the others, more dispersed and more distant from the diagonal $\varphi^+=\varphi$.
We have already called attention to the possible importance of the three reflections in the degree of approximation verified in the angles $\varphi$ and $\varphi^+$. It occurs to us, therefore, to put the following question: is it not precisely the number of reflections that, on differentiating themselves from the $3$ occurrences, interfere so negatively with the behavior of the cavity? The investigations that follow will demonstrate, among other things, that our suspicion on this point has a basis.

The following theorem says that for $\varphi$ outside some interval $(-\varphi_0,\varphi_0)$, the number of reflections is always three. The proof is presented in appendix~\ref{cha:condSuf3col}.
\begin{thm}
\label{teor:3ref}
For $\varphi$ entry angles superior (in absolute value) to $\varphi_0=\arctan\left(\frac{\sqrt{2}}{4}\right)\simeq 19.47^\circ$, the number of reflections to which the particle is subjected in the interior of the Double Parabola cavity is always equal to three, and they occur alternately on the left and right faces of the cavity, no matter what the entry position may be.
\end{thm}

\begin{figure}[!hb]
\begin{center}
\includegraphics*[width=0.4\columnwidth]{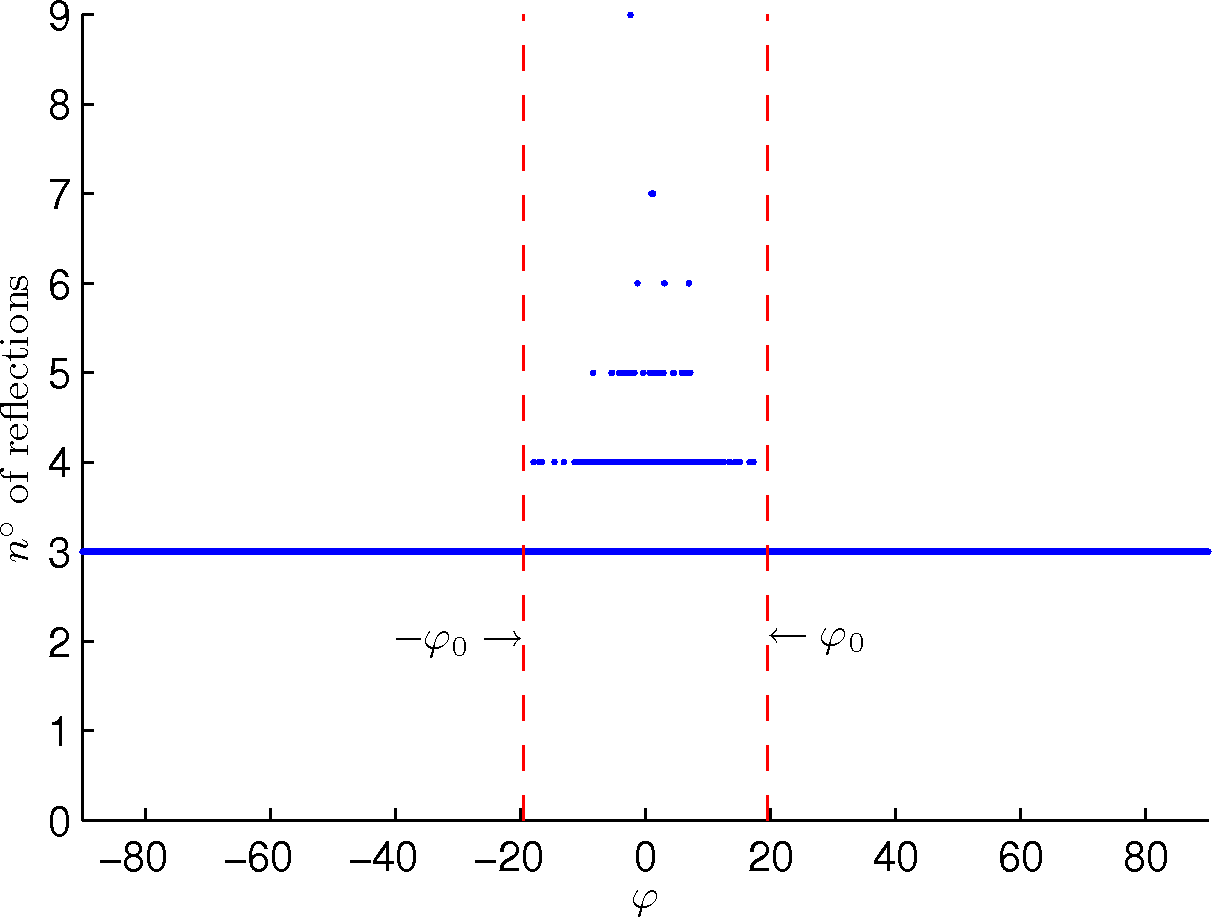}
\caption[Distribution of the $(\varphi,nr)$ pairs on the Cartesian plane.]{Distribution of the $(\varphi,nr)$ pairs on the Cartesian plane, being $nr$ the nº of reflections.}
\label{fig:ncolisoes4}
\end{center}
\end{figure}

As a way to verify that the deductions which we have made are effectively in concordance with the numerical results of the computational model which was developed, we present one more graph, figure~\ref{fig:ncolisoes4}, produced with $10.000$ pairs of $(x,\varphi)$ values, generated randomly with uniform distribution.
%%%%%%%%%%%%%%%%%%%%%%%%%%%%%%%%%%%%%%%%%%%%%%%%%%%%
%%%%%%%%% Figura %%%%%%%%%%%%%%%%%%%%%%%%%%%%%%%%%%%%%%%%%%%%%%%%%%%%%%%%%%%%%%
As can be observed in figure~\ref{fig:ncolisoes4}, all the trajectories with $4$ or more reflections, among the $10.000$ considered, happened within the interval $(-\varphi_0,\varphi_0)$. Outside this interval (for $\left|\varphi\right|>\varphi_0$) the trajectories are always of three reflections. Additionally, we can verify that there isn't any trajectory with less than three reflections. 
%We will prove in appendix~B this to be also a characteristic of the cavity.
This numerical evidence is confirmed by the following theorem:
\begin{thm}
\label{teor:min3ref}
Any particle which enters in the cavity Double Parabola describes a trajectory with a minimum of $3$ reflections.
\end{thm}

The proof of theorem 2 is presented in appendix~\ref{cha:min3col}.

Of the conclusions which we arrived at we can immediately come to the following corollary: in trajectories with $4$ or more reflections the angular difference $\left|\varphi-\varphi^+\right|$, no matter how much bigger it may be, will never be superior to $2\varphi_0\simeq 38.94^\circ$, a value which is much more inferior to the greatest angle which it is possible to form between two vectors ($180^\circ$). The proof of this corollary is simple:
as a trajectory of $4$ or more reflections is always associated with a entry angle $-\varphi_0<\varphi<\varphi_0$, the exit angle will be situated necessarily in the same interval; taking into account the property of reversibility associated with the law of reflection which governs reflections, if just to be absurd we were to admit $\left|\varphi^+\right|>\varphi_0$, on inverting the direction of the particle, we would be in the position of having a trajectory of more than $3$ reflections with a $\varphi^+$ entry angle situated outside the interval $(-\varphi_0,\varphi_0)$, which would enter into contradiction with the initial postulate.

Summarizing:
\begin{itemize}
\item There is verified a great predominance of trajectories with $3$ reflections;
\item There are no trajectories of fewer than $3$ reflections;
\item The critical angle $\varphi_0$ has the value $\varphi_0=\arctan\left(\frac{\sqrt{2}}{4}\right)\simeq 19.47^\circ$;
\item Outside the interval $(-\varphi_0,\varphi_0)$, all the trajectories are of $3$ reflections;
\item In trajectories with $4$ or more reflections, the angular difference is delimited by $2\varphi_0$: $\left|\varphi-\varphi^+\right|<2\varphi_0$.
\end{itemize}

\section{Conclusion and future perspectives}
\label{sec:concl}
\begin{figure}[!ht]
\begin{center}
\includegraphics[width=0.25\columnwidth]{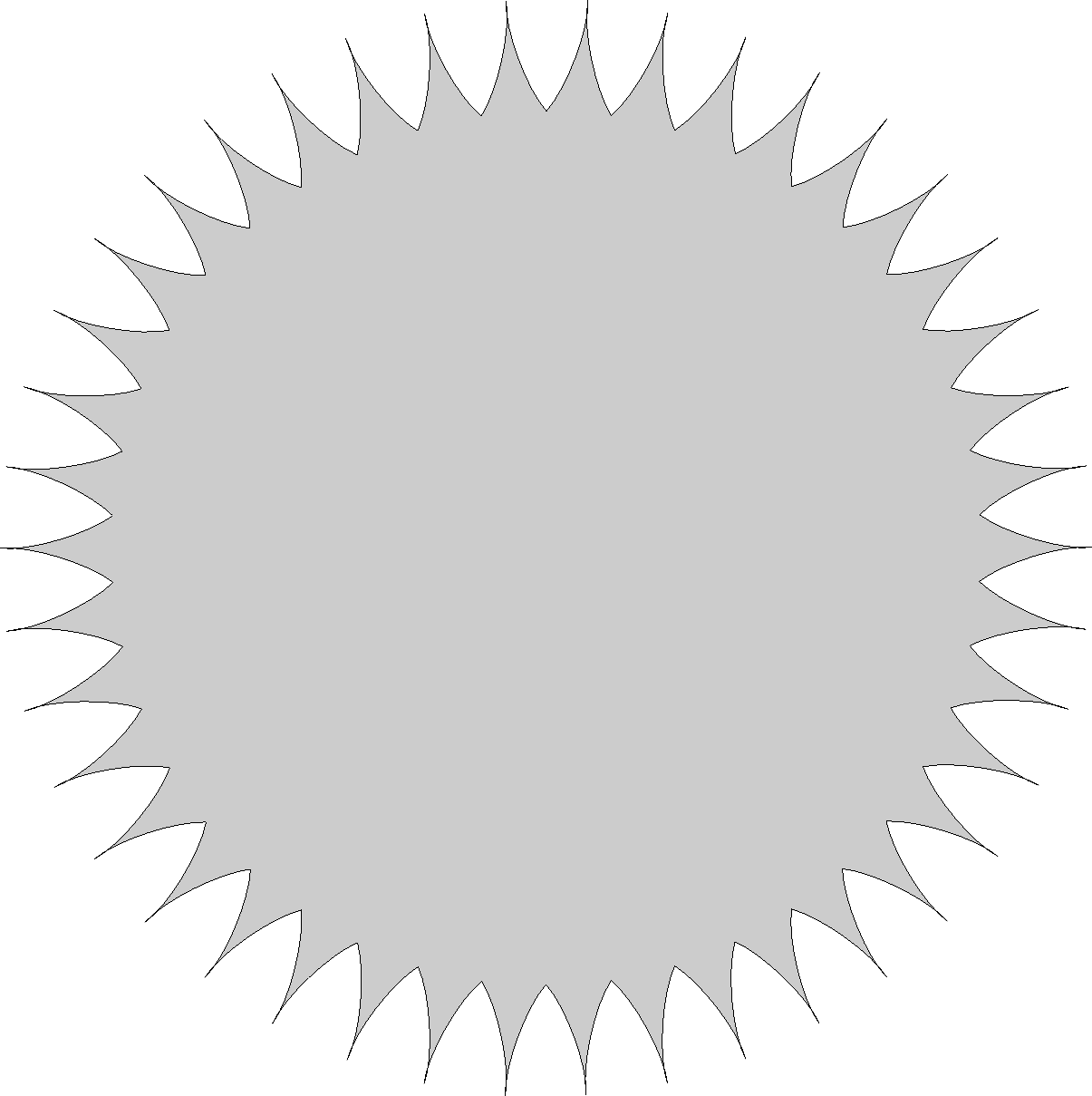}
\caption{(almost) Optimal 2D body.}
\label{fig:discoOptimo}
\end{center}
\end{figure}

In the continuation of the study carried out previously by the authors in \cite{Plakhov07:CM, Plakhov07}, with the work now presented it has been possible to obtain an original result which appears to us to have great scope: the algorithms of optimization converged for a geometrical shape very close to the ideal shape --- the \emph{Double Parabola}. This concerns a form of roughness which confers a nearly maximal resistance (very close to the theoretical upper bound) to a disc which, not only travels in a translational movement but also rotates slowly around itself. In figure~\ref{fig:discoOptimo} one of these bodies is shown.
%%%%%%%%% figura %%%%%%%%%%%%%%%%%%%%%%%%%%%%%%%%%%%%%%%%%%%%%%%%%%%%%%%%%%%%%%
%%%%%%%%% figura %%%%%%%%%%%%%%%%%%%%%%%%%%%%%%%%%%%%%%%%%%%%%%%%%%%%%%%%%%%%%%
Noting that the contour of the presented body is integrally formed by $42$ cavities $\Omega$ with the shape of a Double Parabola, each one of which with a relative resistance of $1.49650$, from \eqref{eq:RB2} and \eqref{eq:RxPerimetros} we conclude that $R(B)=\frac{\sin(\pi/42)}{\pi/42}R(\Omega)\approx 1.4951$ is the total resistance of the body, a value $49.51\%$ above the value of resistance of the corresponding disc of smooth contour (the smallest disc which includes the body). We know that if the body were formed by a sufficiently elevated number of these cavities, its resistance would even reach the value $1.4965$, but the example presented is sufficient in order for us to understand how close we are to the known theoretical upper bound ($50\%$).

Although the value of resistance of the Double Parabola had been determined numerically, an analytical study was done in section~\ref{sec:caract}, with the objective of consolidating the presented results.
We have managed to prove some important properties which help in the understanding of the elevated value of resistance which was obtained. We will try in the future to develop other theoretical studies which will allow us to consolidate this result even further. For example, an interesting open problem lies in delimiting the lag between the angles of entrance and exit for the trajectories of $3$ reflections --- for the others (trajectories with $4$ or more reflections) we already know that $\left|\varphi-\varphi^+\right|<2\varphi_0 \simeq 2 \times 19.47^\circ$.

The Double Parabola is effectively a result of great practical scope.
Besides maximizing Newtonian resistance, it is exciting to verify that the potentialities of the Double Parabola shape found by us could also reveal themselves to be very interesting in other areas of practical interest.
If we coat the interior part of the Double Parabola cavity with a polished ``surface'', the trajectory of the light in its interior will be described by the principles of geometrical optics, in particular rectilinear propagation of light, laws of reflection and reversibility of light. Thus, as the computational models which were developed by us to simulate the dynamic of billiards in the interior of each one of the shapes studied (where collisions of particles are considered perfectly elastic) are equally valid when the problem becomes of an optical nature, we can also look at $2$D shape found by us in this new perspective.
Given the characteristics of reflection which the Double Parabola shape presents we can rapidly conceive for it a natural propensity for being able to be used with success in the design of retroreflectors --- see in \cite{gouvPhD} the exploratory study of its possible utilization in roadway signalization and the automobile industry.

An incursion into the three-dimensional case, carried out in \cite{gouvPhD}, also showed that the Double Parabola is a shape of cavity which is very special. Our conviction of its effectiveness was strongly reinforced when we obtained the best result for the 3D case. This result was achieved with a cavity whose surface is the area swept by the movement of the Double Parabola curve in the direction perpendicular to its plane. The value of its resistance ($R=1.80$) having been a little below the theoretical upper bound for the 3D case ($R=2$), to go beyond this value will be also an interesting challenge to consider in the future.

For the 2D case we envision greater difficulty in going beyond the result which has already been reached --- whether for the proximity which it has to the theoretical upper bound, or for the fact that we have already carried out, without success, a series of investigation with just this objective.

\appendix
\section{Proof of theorem~\ref{teor:3ref}}
\label{cha:condSuf3col}

\begin{figure}[!hb]
\begin{center}
\includegraphics*[width=0.6\columnwidth]{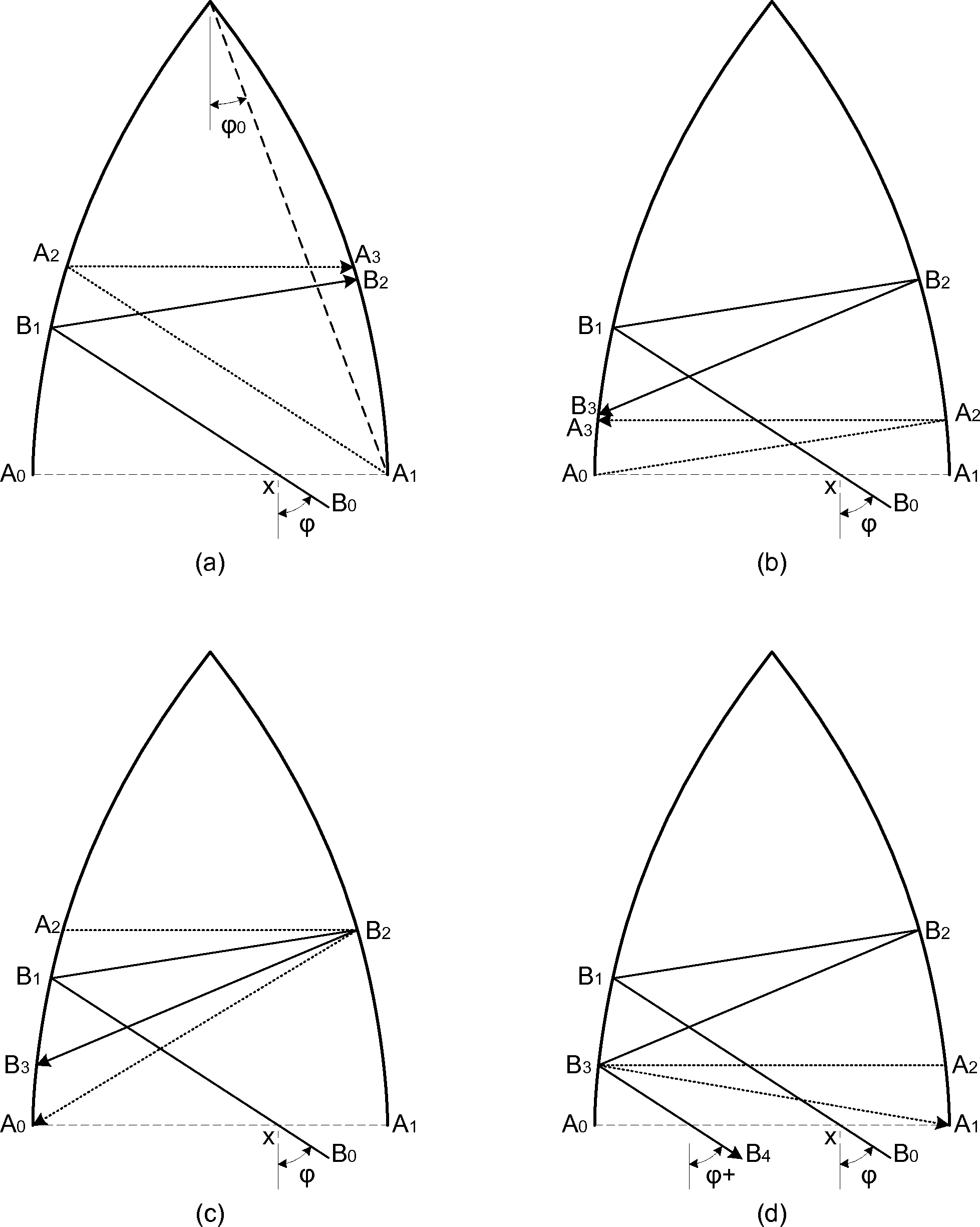}
\caption[Illustrations to the study of the trajectory with entry angle $\varphi>\varphi_0 \simeq 19.47^\circ$.]{Set of illustrations to the study of the trajectory of particles with entry angle $\varphi>\varphi_0 \simeq 19.47^\circ$,
in the cavity ``Double Parabola''.}
\label{fig:parabOpt3col}
\end{center}
\end{figure}

Consider a particle which enters into the cavity in $(x,0)$, with the vector velocity forming an angle $\varphi$ with the vertical axis, just as is found represented in the illustrations of figure~\ref{fig:parabOpt3col}, where we assume that the axis of symmetry of the cavity is the axis of the $y$ and that its base $\overline{A_0A_1}$ is placed on the axis $x$. In this way, the position of the particle at entry of the cavity assumes only values in the interval $\left(-\frac{1}{2},\,\frac{1}{2}\right)\times\{0\}$.
%%%%%%%%% Figura %%%%%%%%%%%%%%%%%%%%%%%%%%%%%%%%%%%%%%%%%%%%%%%%%%%%%%%%%%%%%%
%%%%%%%%% Figura %%%%%%%%%%%%%%%%%%%%%%%%%%%%%%%%%%%%%%%%%%%%%%%%%%%%%%%%%%%%%%
%%%%%%%%% Figura %%%%%%%%%%%%%%%%%%%%%%%%%%%%%%%%%%%%%%%%%%%%%%%%%%%%%%%%%%%%%%
%%%%%%%%% Figura %%%%%%%%%%%%%%%%%%%%%%%%%%%%%%%%%%%%%%%%%%%%%%%%%%%%%%%%%%%%%%

Given the symmetry of the cavity in relation to its vertical axis, it will be enough to analyze its behavior for $\varphi_0 <\varphi< 90^\circ$. The conclusions at which we arrive will be in this way equally valid for $-90^\circ<\varphi< -\varphi_0$.

We will analyze therefore in detail and separately each one of the sub-trajectories which compose all the trajectory described by the movement of the particle in the interior of the cavity.

%\vspace{0.3cm}
\noindent
\emph{Sub-trajectory $\overrightarrow{B_0B_1}$}

For $\varphi>\varphi_0$, we have the guarantee that the first reflection occurs in the parabolic curve of the left side of the cavity, just as can be easily deduced from the illustration (a). So that the particle collides with the left curve it is enough that the $\varphi$ angle is superior to $\arctan(x/\sqrt{2})$, a magnitude which has as upper bound $\varphi_0=\arctan(\sqrt{2}/4)$. We thus have the initial trajectory of the cavity represented in illustration (a) by vector $\overrightarrow{B_0B_1}$.

%\vspace{0.3cm}
\noindent
\emph{Sub-trajectory $\overrightarrow{B_1B_2}$}
\label{pg:subtrajB1B2}

After colliding in $B_1$, in agreement with the law of reflection, the particle follows trajectory $\overrightarrow{B_1B_2}$. We prove that $\overrightarrow{B_1B_2}$ has an ascendant path --- illustration (a). We trace the straight line $\overline{A_1A_2}$, segment, parallel to the initial trajectory of the particle $\overline{B_0B_1}$, which passes through the focus of the left parabola ($A_1$). Because of the focal property of this parabola, a particle which takes the sub-trajectory $\overrightarrow{A_1A_2}$, after reflection at $A_2$, will follow a horizontal direction $\overline{A_2A_3}$ (proceeding after its trajectory, after a new reflection, in the direction of the focus $A_0$ of the second parabola).
Upon the occurrence of the first reflection of the particle at $B_1$, a point of the curve necessarily positioned below $A_2$, the trajectory $\overrightarrow{B_1B_2}$, which it will follow straight away, will be on an ascendant path, since the derivative $\frac{\mathrm{d} y}{\mathrm{d} x}$ of the curve at this point ($B_1$) is superior to the derivative in $A_2$, where the trajectory followed was horizontal.

\begin{figure}[!hb]
\begin{center}
\includegraphics*[width=0.57\columnwidth]{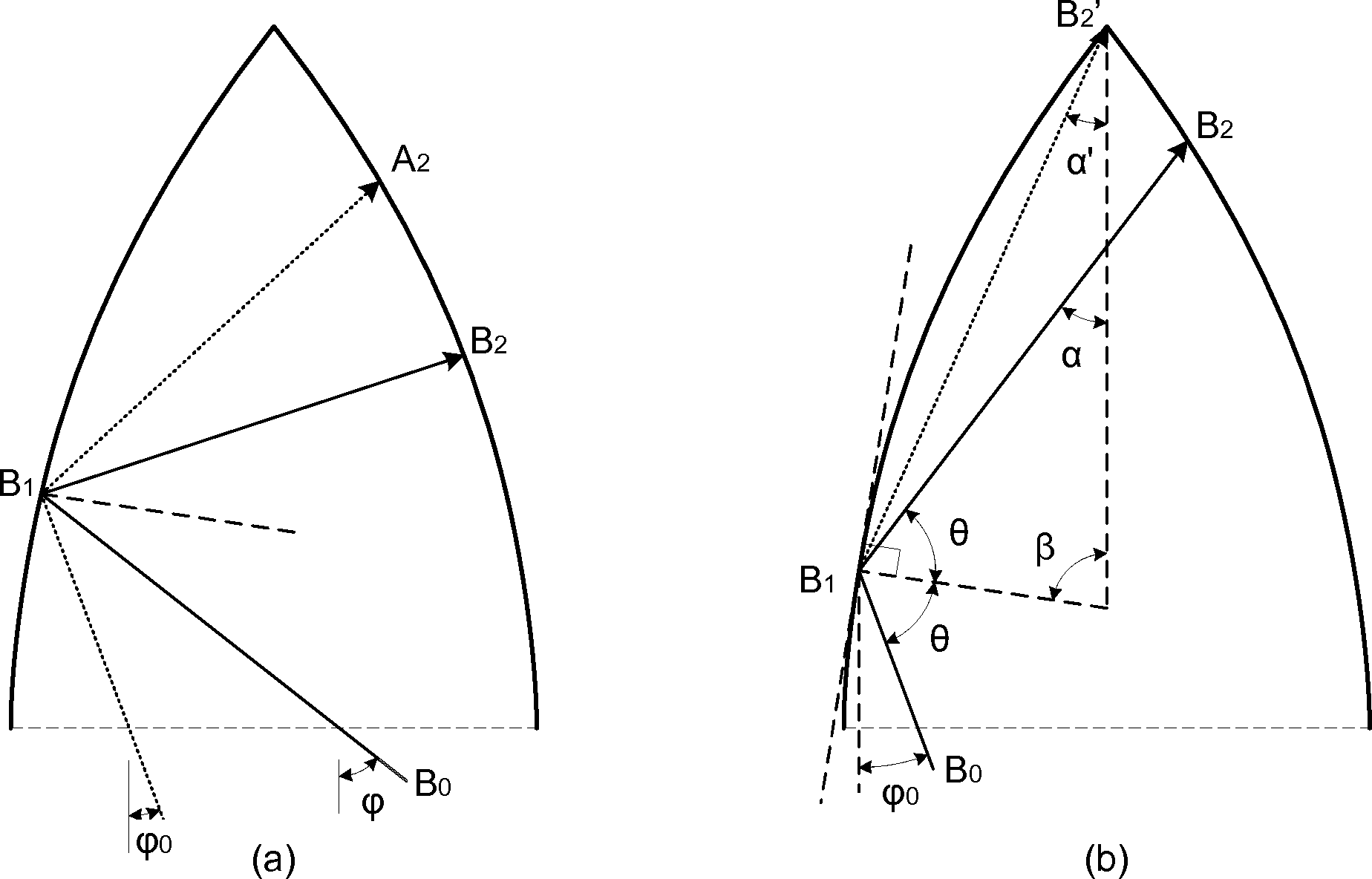}
\caption{Illustrations to the study of the second reflection.}
\label{fig:parabOpt3colA}
\end{center}
\end{figure}

Although we now know that $\overrightarrow{B_1B_2}$ takes an ascendant path, nothing yet guarantees to us that the second reflection happens necessarily in the parabola of the right side. If we are able to verify that for $\varphi=\varphi_0$ the second reflection is always on the right side, no what the entry position $x$ is, therefore, logically, the same will happen for any value $\varphi>\varphi_0$.
This premise can be easily accepted with the help of illustration (a) of figure~\ref{fig:parabOpt3colA}: for any value of $\varphi>\varphi_0$, with the first reflection at a given point $B_1$, it is always possible to trace a trajectory for $\varphi=\varphi_0$ which presents the first reflection at the same point $B_1$;
the second reflection at the curve of the right side being for the case  $\varphi=\varphi_0$, necessarily the same will happen for the trajectory with $\varphi>\varphi_0$, since the angle of reflection will be less in this second case, just as is illustrated in the figure.
Consequently, it will be enough for us to prove for $\varphi=\varphi_0$, that the second reflection always occurs in the parabola of the right side, so that the same is proven for any which is the $\varphi>\varphi_0$.
%%%%%%%%% Figura %%%%%%%%%%%%%%%%%%%%%%%%%%%%%%%%%%%%%%%%%%%%%%%%%%%%%%%%%%%%%%
%%%%%%%%% Figura %%%%%%%%%%%%%%%%%%%%%%%%%%%%%%%%%%%%%%%%%%%%%%%%%%%%%%%%%%%%%%
%%%%%%%%% Figura %%%%%%%%%%%%%%%%%%%%%%%%%%%%%%%%%%%%%%%%%%%%%%%%%%%%%%%%%%%%%%
%%%%%%%%% Figura %%%%%%%%%%%%%%%%%%%%%%%%%%%%%%%%%%%%%%%%%%%%%%%%%%%%%%%%%%%%%%

In illustration (b) of figure~\ref{fig:parabOpt3colA} the trajectory until the second reflection of a particle with the angle of entry $\varphi_0$
($\overrightarrow{B_0B_1}$ and $\overrightarrow{B_1B_2}$) is shown.
As one can conclude from the illustration, the $B_2$ reflection only will happen on the curve of the left side if the $\alpha$ angle is less than $\alpha'$. We have determined the value of the two angles.

Being $(x_1,y_1)$ the coordinates of the point $B_1$, we will have $\tan(\alpha')=-x_1/(\sqrt{2}-y_1)$, thus
\begin{equation}
\label{eq:alphalinha}
\alpha'
=\arctan\left(\frac{-(y_1^2/4-1/2)}{\sqrt{2}-y_1}\right)
=\arctan\left(\frac{(2-y_1^2)/4}{\sqrt{2}-y_1}\right)
=\arctan\left(\frac{\sqrt{2}+y_1}{4}\right)\text{.}
\end{equation}
In order to arrive at the value of $\alpha$ we resolve the system of three equations, of unknown $\alpha$, $\theta$ and $\beta$, which are taken directly from the geometry of the actual figure
\begin{equation*}
\left\{
\begin{array}{l}
\alpha + \beta + \theta = \pi\\
\beta = \varphi_0+\theta\\
\arctan \left(\frac{1}{2}y_1\right)+\varphi_0+\theta = \frac{\pi}{2}
\end{array}
\right.
\end{equation*}
The tangent line to the curve in $B_1$ makes with the vertical an angle whose tangent has as its value the derivative $\frac{\mathrm{d} x}{\mathrm{d} y}$ of the curve at that point (in $y=y_1$), where $\frac{\mathrm{d} x}{\mathrm{d} y}
=\frac{\mathrm{d}}{\mathrm{d} y}(\frac{1}{4}y^2-\frac{1}{2})=\frac{1}{2}y$. Because of this, that angle emerges represented in the third of the equations by the magnitude $\arctan(y_1/2)$.
By resolving the system, the following result is obtained for $\alpha$
\begin{equation}
\label{eq:alpha}
\alpha = \varphi_0+2\arctan({y_1}/{2})=\arctan({\sqrt{2}}/{4})+2\arctan({y_1}/{2})
\text{.}
\end{equation}
Finally we prove that $\alpha>\alpha'$, no matter what $y_1\in \, (0,\sqrt{2})$ is.
Of the equations~\eqref{eq:alpha} and \eqref{eq:alphalinha}, it will be equivalent to proving
%\begin{eqnarray*}
$$\arctan(\sqrt{2}/4)+2\arctan(y_1/2)>\arctan((\sqrt{2}+y_1)/4)\text{.}$$
%\end{eqnarray*}
Given that $0<y_1<\sqrt{2}$, both of the members of the inequality represent angles situated in the first quadrant of the trigonometrical circle. Because of this we can maintain the inequality for the tangent of the respective angles. Applying the tangent to both of the members, after effecting some trigonometrical simplifications, we arrive at the following relation
\begin{equation*}
\frac{1}{4}\,\frac{4\sqrt{2}-\sqrt{2}y_1^2+16 y_1}{4-y_1^2-\sqrt{2} y_1}>
\frac{\sqrt{2}+y_1}{4} 
\end{equation*}
Which, with additional algebraic simplifications, takes the form
\begin{equation*}
{y_1\left(\sqrt{2}y_1+14+y_1^2\right)}/({4-y_1^2-\sqrt{2} y_1})>0 \text{.}
\end{equation*}
As $0<y_1<\sqrt{2}$, it can easily be seen that both the numerator and the denominator of the fraction present in this latest inequality are positive magnitudes. Thus $\alpha>\alpha'$, which contradicts the condition which was necessary so that the reflection $B_2$ could occur on the curve of the left side, it therefore being proven, as we intended, that in no situation does the reflection $B_2$ of illustration (b) of figure~\ref{fig:parabOpt3colA} happen on the curve of the left side. Logically, we can therefore conclude that the same happens for whatever is $\varphi>\varphi_0$: the second reflection of the particle occurs always in the parabola of the right side.

%\vspace{0.3cm}
\noindent
\emph{Sub-trajectory $\overrightarrow{B_2B_3}$}
\label{pg:subtrajB2B3}

We prove that the sub-trajectory $\overrightarrow{B_2B_3}$ has a descendant path --- illustration (b) of figure~\ref{fig:parabOpt3col}. Imagine, for this purpose, a sub-trajectory $\overrightarrow{A_0A_2}$, parallel to $\overline{B_1B_2}$ and which passes through the focus $A_0$.
The sub-trajectory $\overrightarrow{A_2A_3}$ which will follow the reflection in $A_2$ --- a point of the parabola on the right side situated below $B_2$ --- will be horizontal. The derivative of the curve in $A_2$ being superior to the derivative value in $B_2$, the sub-trajectory $\overline{B_2B_3}$ will necessarily be of a descending nature.

Even if we already know that the sub-trajectory is descendant, we have not yet shown that sub-trajectory in no situation conducts the particle directly to the exit of the cavity. Therefore follows the proof that the reflection $B_3$ always occurs in a position superior to $A_0$ --- illustration (c) of figure~\ref{fig:parabOpt3col}.
We trace $\overline{A_2B_2}$, a segment of the horizontal straight line which passes through the point of reflection $B_2$. If the particle followed this trajectory, it would collide at the same point $B_2$, but heads itself to $A_0$. Therefore, by the law of reflection, $B_3$ will have to be above $A_0$, since $\overline{B_1B_2}$ makes an angle with the normal vector at the curve in $B_2$ less than that formed by segment $\overline{A_2B_2}$.

%\vspace{0.3cm}
\noindent
\emph{Sub-trajectory $\overrightarrow{B_3B_4}$}

We will now show that the sub-trajectory which follows the reflection at $B_3$ crosses the segment $\overline{A_0A_1}$, that is, directs itself to the outside of the cavity --- illustration (d) of figure~\ref{fig:parabOpt3col}.
We trace, therefore, $\overline{A_2B_3}$, a segment of horizontal straight line which passes through the point of reflection $B_3$. If the particle followed this trajectory, it would collide at $B_3$ and would head itself towards $A_1$.
Therefore, by the law of reflection, the straight line where the sub-trajectory $\overrightarrow{B_3B_4}$ is placed will have necessarily to pass below $A_1$,
since $\overline{B_2B_3}$ makes an angle with the normal vector at the curve in $B_3$ bigger than that formed by the segment $\overline{A_2B_3}$.
We have shown that the sub-trajectory crosses the axis of the $x$ at a point situated to the left of $A_1$, but we have not yet shown that it occurs to the right of $A_0$. For that, we will have to prove that the third is the last of the reflections, that is, that in no situation does there occur a fourth reflection in the parabola of the left side. There follows this proof, of them all the most complex one.

\begin{figure}[!hb]
\begin{center}
\includegraphics*[width=0.57\columnwidth]{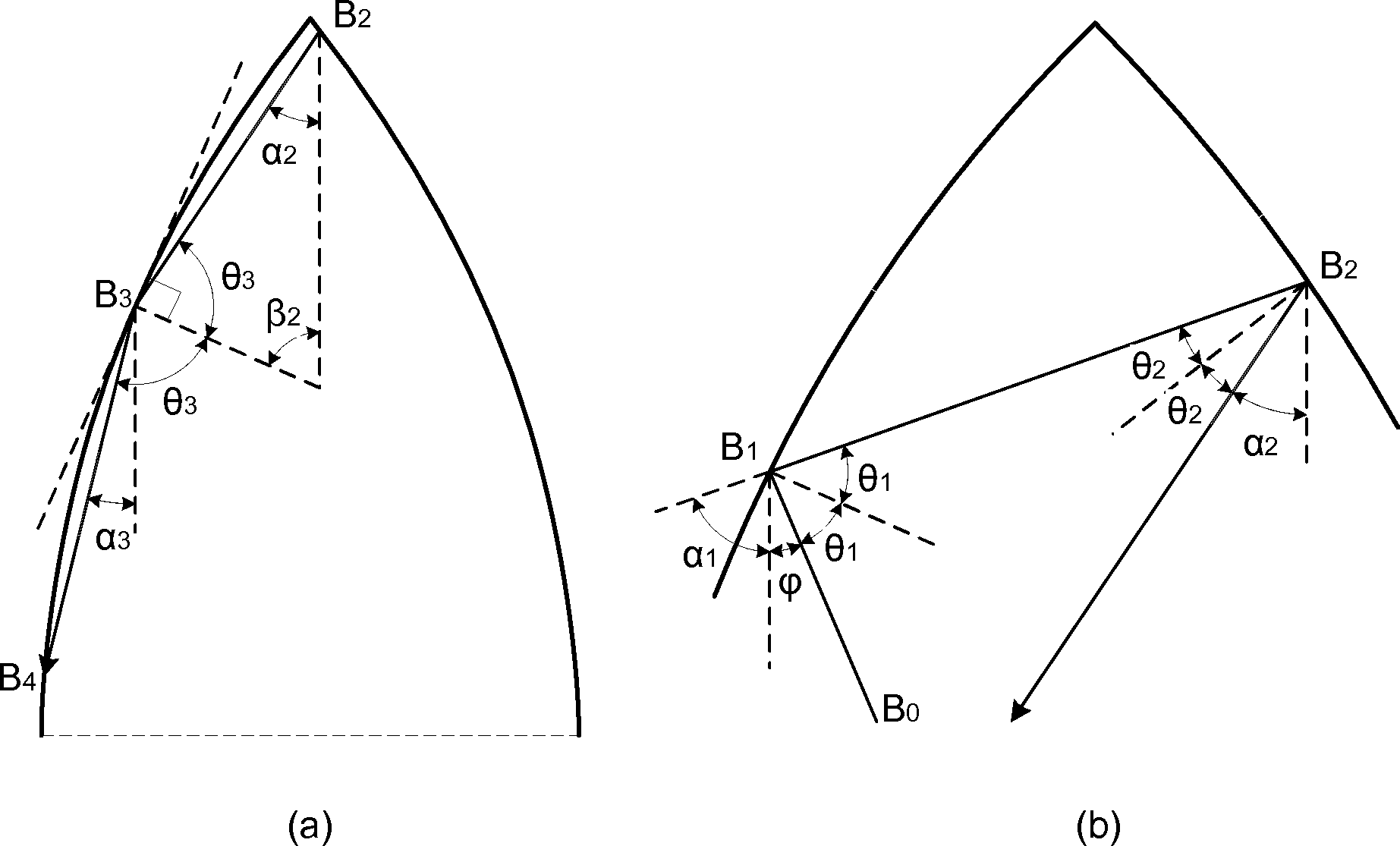}
\caption{Illustrations to the study of a hypothetical fourth reflection.}
\label{fig:parabOpt3colB}
\end{center}
\end{figure}

In order to prove that following the third reflection there occurs no other collision in the left parabola, we will show that a fourth collision --- represented by $B_4$ in the illustration (a) of figure~\ref{fig:parabOpt3colB} --- has its origin always in an entry angle $\varphi$ inferior to $\varphi_0$.
%%%%%%%%% Figura %%%%%%%%%%%%%%%%%%%%%%%%%%%%%%%%%%%%%%%%%%%%%%%%%%%%%%%%%%%%%%
%%%%%%%%% Figura %%%%%%%%%%%%%%%%%%%%%%%%%%%%%%%%%%%%%%%%%%%%%%%%%%%%%%%%%%%%%%
%%%%%%%%% Figura %%%%%%%%%%%%%%%%%%%%%%%%%%%%%%%%%%%%%%%%%%%%%%%%%%%%%%%%%%%%%%
%%%%%%%%% Figura %%%%%%%%%%%%%%%%%%%%%%%%%%%%%%%%%%%%%%%%%%%%%%%%%%%%%%%%%%%%%%
We will thus study the trajectory of the particle in the inverse order of its progression: we commence by admitting the existence of the sub-trajectory $\overrightarrow{B_3B_4}$ of illustration (a) and we will analyze its implications in all the preceding trajectory.

In illustration (a) of figure~\ref{fig:parabOpt3colB} are to be found represented the sub-trajectories $\overrightarrow{B_2B_3}$ and $\overrightarrow{B_3B_4}$.
We begin by relating $\alpha_2$ with $\alpha_3$, the angles which the vectors $\overrightarrow{B_2B_3}$ and $\overrightarrow{B_3B_4}$, respectively, form with the vertical axis. For these purposes we resolved the system of three equations, of unknown $\alpha_2$, $\theta_3$ and $\beta_2$, which are taken from the geometry of the figure,\footnote{The variables denoted by $x_i$ and $y_i$, with $i =1,\ldots,4$, represent the coordinates of $i$-th point of reflection, identified by $B_i$.}
\begin{equation*}
\left\{
\begin{array}{l}
\theta_3 =\beta_2 + \alpha_3\\
\arctan \left(\frac{1}{2}y_3\right)+\beta_2 = \frac{\pi}{2}\\
\alpha_2 +\theta_3+\beta_2 = \pi
\end{array}
\right.
\end{equation*}
obtaining
\begin{equation*}
\label{eq:alpha2}
\alpha_2 =2\arctan({y_3}/{2})-\alpha_3
\text{,}
\end{equation*}
in which $\arctan(y_3/2)$ is the angle which the straight line tangent at the curve in $B_3$ makes with the vertical --- the inclination of the straight line tangent is given by $\frac{\mathrm{d} x}{\mathrm{d} y}\left.\right|_{y=y_3}
=\frac{\mathrm{d}}{\mathrm{d} y}(\frac{1}{4}y^2-\frac{1}{2})\left.\right|_{y=y_3}
=\frac{1}{2}y_3$.
In its turn, the angle $\alpha_3$ can be expressed in the following way:
%\begin{equation*}
%\label{eq:alpha3}
$\alpha_3 =\arctan(({x_3-x_4})/({y_3-y_4}))
=\arctan(({\frac{1}{4}y_3^2-\frac{1}{4}y_4^2})/({y_3-y_4}))
=\arctan({(y_3+y_4)}/{4})\text{,}$
%\end{equation*}
which permits us to write $\alpha_2$ in function only of the ordinates $y_3$ and $y_4$ of the extremes of the vector $\overrightarrow{B_3B_4}$,
\begin{equation}
\label{eq:alpha2b}
\alpha_2 =2\arctan ({y_3}/{2})-\arctan(({y_3+y_4})/{4})
\text{.}
\end{equation}

In order to be able to prove what we intend --- impossibility of occurrence of the reflection $B_4$ --- we need to find a lower bound for the ordinate of the position where each one of the four reflections occurs, or in other words, to determine $\left\{y_1^*,y_2^*,y_3^*,y_4^*\right\}$, just that
\begin{equation}
\label{eq:minorantes}
y_1\geq y_1^*,\;y_2\geq y_2^*,\;y_3\geq y_3^*,\;y_4\geq y_4^*,\;
\;\forall (\varphi,x)\in(\varphi_0,{\pi}/{2})\times (-{1}/{2},{1}/{2})\text{.}
\end{equation}
It can easily be understood that $y_4^*=0$. We will therefore determine the other three lower bounds, commencing with $y_2^*$.

We know that $0<y_4<y_3$; therefore, from \eqref{eq:alpha2b} we take away that
\begin{equation*}
\label{eq:alpha2Limites}
\arctan({y_3}/{2})<\alpha_2 < 2\arctan ({y_3}/{2})-\arctan ({y_3}/{4})
\text{.}
\end{equation*}
Being aware that $\alpha_2$ is situated in the first quadrant of the trigonometrical circle, we can maintain the inequalities for the tangent of the respective angles. After some algebraic simplifications, we obtain
\begin{equation}
\label{eq:tanAlpha2Limites}
{y_3}/{2}<\tan(\alpha_2) < {y_3\left(12+y_3^2\right)}/{16}
\text{.}
\end{equation}
The equation of the straight line which connects $B_2$ to $B_3$ takes the form
%\begin{equation*}
$x=m(y-y_3)+x_3,$
%\end{equation*}
with $m=\tan(\alpha_2)$ and $x_3=\frac{1}{4}y_3^2-\frac{1}{2}$. As we are interested in finding the ordinate of the point of interception of this straight line with the parabolic curve situated on the right side, with equation
%\begin{equation*}
$x=-\frac{1}{4}y^2+\frac{1}{2}
\text{,}$
%\end{equation*}
we have to resolve the equation of second degree, in the variable $y$, which results in the elimination of the variable $x$ by combination of the two previous equations. The ordinate $y_2$, of the second reflection, being the positive root of the equation, takes the form
%\begin{equation*}
$y_2 = -2 m+\sqrt{4 m^2+4 m y_3 - y_3^2+4}
\text{ .}$
%\end{equation*}

The magnitude $y_2$ is expressed in function of two variables, $m$ and $y_3$, which as we know assume only positive values. So as to accept more easily the deductions which we are going to make in the tracking of $y_2^*$, we will imagine, without any loss of generality, that $y_3$ is a fixed value. We begin by showing that the derivative of $y_2$ in order to the variable $m$,
\begin{equation}
\label{eq:derivaday2}
\frac{\mathrm{d} y_2}{\mathrm{d} m}=
\frac{2\left(2m+y_3 -\sqrt{4 m^2+4m y_3 -y_3^2 + 4}\right)}{\sqrt{4 m^2+4m y_3 -y_3^2 + 4}}
\text{ ,}
\end{equation}
has a negative value for whatever value of $y_3$ is. As $y_3<\sqrt{2}$, inevitably $y_3^2<4$, thus the two radicands $(4m^2+4my_3-y_3^2+4)$ present in the equation~\eqref{eq:derivaday2} have always a positive value.
The restriction $y_3<\sqrt{2}$ allows us still to successively deduce the following inequalities
 \vspace*{-3mm}

\begin{eqnarray*}
y_3^2<2 \Leftrightarrow 2 y_3^2<4 \Leftrightarrow y_3^2<4-y_3^2
\Leftrightarrow 4 m^2+4 m y_3 + y_3^2<4 m^2+4 m y_3 +4-y_3^2\\
\Leftrightarrow  \left(2 m+ y_3\right)^2<4 m^2+4 m y_3 -y_3^2+4
\Leftrightarrow 2 m+ y_3<\sqrt{4 m^2+4 m y_3 -y_3^2+4}
\text{ .}
\end{eqnarray*}
This last inequality confirms that $\frac{\mathrm{d} y_2}{\mathrm{d} m}<0$, for whatever $y_3$ may be. In this way, the value $y_2$ is so much less the greater is the value of $m$. As is examined in \eqref{eq:tanAlpha2Limites}, $m<M={y_3\left(12+y_3^2\right)}/{16}$, thus
%\begin{equation*}
$y_2 > -2 M+\sqrt{4 M^2+4 M y_3 - y_3^2+4}
\text{ .}$
%\end{equation*}
Substituting $M$, there is obtained, after some simplifications,
\begin{equation}
\label{eq:y2minor}
y_2 > f(y_3), \text{ with } f(y_3)=-\frac{3}{2} y_3-\frac{1}{8} y_3^3+\frac{1}{8} \sqrt{272 y_3^2+40 y_3^4+y_3^6+256}
\text{ .}
\end{equation}
In order to find the minimum value of $f(y_3)$ we begin by computing its derivative:
\begin{equation*}
\frac{\mathrm{d}}{\mathrm{d} y_3}f(y_3)= \frac{272 y_3+80 y_3^3+3 y_3^5-(12+3 y_3^2) \sqrt{272 y_3^2+40 y_3^4+y_3^6+256}}
{8 \sqrt{272 y_3^2+40 y_3^4+y_3^6+256}}
\text{.}
\end{equation*}
The radicands being clearly positive, we only have to concern ourselves with the numerator of the fraction. To find the roots of the derivative function $\frac{\mathrm{d}}{\mathrm{d} y_3}f(y_3)$  is equivalent because of this to resolving the equation
\begin{equation*}
\left(272 y_3+80 y_3^3+3 y_3^5\right)^2=\left(12+3 y_3^2\right)^2 \left(272 y_3^2+40 y_3^4+y_3^6+256\right)
\text{,}
\end{equation*}
which can be simplified in the following:
\begin{equation*}
2304-1024 y_3^2-992 y_3^4-160 y_3^6-3 y_3^8=0 \text{.}
\end{equation*}
This polynomial equation has only one real positive root, of the value
%\begin{equation*}
$\tilde{y}_3=\frac{2}{3} \sqrt{-51 + 6 \sqrt{79}}\simeq 1.017 \text{,}$
%\end{equation*}
signifying that $f(y_3)$ has a global minimum in $\tilde{y}_3$, because, as we show in what follows, $\frac{\mathrm{d}^2}{\mathrm{d} y_3^2}f(y_3)>0$ and the function does not presents other points of stationarity.

We thus show that $\frac{\mathrm{d}^2}{\mathrm{d} y_3^2}f(y_3)>0$, with
\vspace{-4mm}

\begin{eqnarray}
\label{eq:y2segDeriv}
\frac{\mathrm{d}^2}{\mathrm{d} y_3^2}f(y_3)= \frac{
18240 y_3^4+2960 y_3^6+180 y_3^8+3 y_3^{10}+34816+30720 y_3^2
}{
4 (272 y_3^2+40 y_3^4+y_3^6+256)^\frac{3}{2}
} \nonumber\\
-\frac{
(816 y_3^3+120 y_3^5+3 y_3^7+768 y_3) \sqrt{272 y_3^2+40 y_3^4+y_3^6+256}
}{
4 (272 y_3^2+40 y_3^4+y_3^6+256)^\frac{3}{2}
}
\text{.}
\end{eqnarray}
We show that $\frac{\mathrm{d}^2}{\mathrm{d} y_3^2}f(y_3)>0$ is equivalent to showing that the numerator of the first fraction is superior to the numerator of the second, in \eqref{eq:y2segDeriv}.
After elevating the two terms to the square, we arrive at the inequality
\begin{eqnarray*}
-3 y_3^{16}+36 y_3^{14}+9464 y_3^{12}+191616 y_3^{10}+1514112 y_3^8+5817344 y_3^6
\nonumber\\
+13535232 y_3^4+15532032 y_3^2+9469952>0
\text{.}
\end{eqnarray*}
We easily prove the veracity of this relation, given that we have a unique negative term ($-3 y_3^{16})$ which, for example, is inferior in absolute value to the constant term ($9469952$),
%\begin{equation*}
~$y_3<\sqrt{2} \Rightarrow 3 y_3^{16}<768<9469952 \text{.}$
%\end{equation*}
The proof is thus complete that $f(y_3)$ has a global minimum in $\tilde{y}_3$, of the value
%\begin{equation*}
$f(\tilde{y}_3)=\frac{8}{9} \sqrt{-51 + 6 \sqrt{79}} \text{.}$
%\end{equation*}

Accordingly, from \eqref{eq:y2minor}, we finally conclude that
\begin{equation}
\label{eq:y2min}
y_2>y_2^*=\frac{8}{9} \sqrt{-51 + 6 \sqrt{79}} \simeq 1.356 \text{.}
\end{equation}
A lower bound is then found for the height of the second reflection $B_2$ (illustration (a) of figure~\ref{fig:parabOpt3colB}).
We now determine $y_3^*$, a lower bound for the height of the third reflection $B_3$.

So that the reflection $B_2$ occurs in the parabola of the right side it is necessary that the angle $\alpha_2$ is greater than the angle formed between the vertical axis and the segment of straight line which unites $B_3$ with the superior vertex of the cavity,
\vspace{-3mm}
\begin{equation}
\label{eq:alpha2Maj}
\alpha_2>
\arctan \Big(\frac{-x_3}{\sqrt{2}-y_3}\Big)
=\arctan \Big(\frac{-\frac{1}{4}y_3^2+\frac{1}{2}}{\sqrt{2}-y_3}\Big)
=\arctan \Big(\frac{\sqrt{2}+y_3}{4}\Big)
\text{.}
\end{equation}
This inequality, in conjunction with the second relation of inequality of \eqref{eq:tanAlpha2Limites}, allows us to write
\begin{equation*}
({\sqrt{2}+y_3})/{4}<\tan(\alpha_2) <
{y_3\left(12+y_3^2\right)}/{16} \Rightarrow
({\sqrt{2}+y_3})/{4} <
{y_3\left(12+y_3^2\right)}/{16}
\text{,}
\end{equation*}
from which results the inequality
\begin{equation}
\label{eq:ineq3g}
y_3^3+ 8y_3+ 4\sqrt{2}>0
\text{.}
\end{equation}

As the polynomial of the left hand-side of~\eqref{eq:ineq3g} has a positive derivative and admits a unique real root, we immediately conclude that it constitutes an inferior limit for $y_3$, this limit being
\begin{equation}
\label{eq:y3min}
y_3^*=\frac{1}{3}\left(54\sqrt{2}+6\sqrt{546}\right)^\frac{1}{3}
-{8}{\left(54\sqrt{2}+6\sqrt{546}\right)^\frac{-1}{3}}
\simeq 0.670 \text{.}
\end{equation}
It is left to us to determine $y_1^*$, a lower bound for the value of $y_1$ --- the ordinate where the first reflection occurs. To this end we resort to illustration (b) of figure~\ref{fig:parabOpt3colB}, which gives us a more detailed representation of the part of the cavity where the first two reflections occur, $B_1$ and $B_2$. The scheme presented was constructed counting that the first reflection ($B_1$) occurs at a point which is more elevated than that of the third reflection ($B_3$). This is in fact the situation. This itself can be proven by showing that $\alpha_2$ is always smaller than the angle formed between the normal vector at the curve in $B_2$ and the vertical axis, or that is
%\begin{equation*}
$\alpha_2<\frac{\pi}{2}-\arctan\left(\frac{1}{2}y_2\right)
\text{.}$
%\end{equation*}

Taking, in \eqref{eq:tanAlpha2Limites}, at the upper limit of $\tan(\alpha_2)$ and having in mind that $y<\sqrt{2}$, we build the following sequence of inequalities which proves what is intended:
\vspace{-2mm}
\begin{equation*}
\alpha_2 < \arctan \left(\frac{y_3\left(12+y_3^2\right)}{16}\right)
< \overbrace{
\arctan \left(\frac{7\sqrt{2}}{8}\right)
}^{\simeq 51.06^\circ}%\nonumber\\
< \overbrace{\frac{\pi}{2}-\arctan\left(\frac{\sqrt{2}}{2}\right)}^{\simeq 54.74^\circ}
<\frac{\pi}{2}-\arctan\left(\frac{1}{2}y_2\right)
\text{.}
\end{equation*}

We now try to find $y_1^*$. We can define $y_1^*$ as being the ordinate of the point of interception of the left parabola with the semi-straight line with its origin at point $B_2$, positioning as low as possible ($y_2=y_2^*$), and with equal slope to the largest value permitted for the slope of the trajectory which preceded $B_2$ ($\overrightarrow{B_1B_2}$). The equation of the straight line which connects $B_1$ to $B_2$ takes the form
%\begin{equation*}
$x=m(y-y_2)+x_2\text{,}$
%\end{equation*}
with $m=\tan(\alpha_1)$ and $x_2=-\frac{1}{4}y_2^2+\frac{1}{2}$. As we are interested in finding the point of intersection of this straight line with the parabolic curve situated on the left side, with equation
%\begin{equation*}
$x=\frac{1}{4}y^2-\frac{1}{2}
\text{ ,}$
%\end{equation*}
we will have to resolve the equation of the second degree, in the variable $y$, which results in the elimination of the variable $x$ by combination of the two previous equations. Although we have two positive real roots, we are only interested in the smaller of the two, which takes the form
\begin{equation}
\label{eq:y1}
y_1 = 2 m-\sqrt{4 m^2-4 m y_2 - y_2^2+4}
\text{ .}
\end{equation}
As we said, if we do $y_2=y_2^*$ and place the maximum slope to the straight line, which in the previous equations is equivalent to considering $m$ minimum, we obtain $y_1 = y_1^*$. Being $m=\tan(\alpha_1)$, we should determine the value of $\alpha_1$ through the system of equations
\begin{equation*}
\left\{
\begin{array}{l}
\alpha_1 =2\theta_2 + \alpha_2\\
\arctan \left(\frac{1}{2}y_2\right)+\alpha_2 +\theta_2 = \frac{\pi}{2}
\end{array}
\right.
\end{equation*}
which is taken from the geometry of illustration (b) of figure~\ref{fig:parabOpt3colB}. Is obtained
\begin{equation}
\label{eq:alpha1}
\alpha_1 =\pi-2\arctan ({y_2}/{2})-\alpha_2
\text{.}
\end{equation}
From this latest equality, from \eqref{eq:tanAlpha2Limites}, and given that $y_2<\sqrt{2}$ and $y_3<\sqrt{2}$, we deduce that
%\begin{equation*}
$\alpha_1 > \pi-2\arctan ({y_2}/{2})-\arctan ({y_3\left(12+y_3^2\right)}/{16})%\nonumber\\
> \pi-2\arctan ({1/\sqrt{2}})-\arctan (7\sqrt{2}/8)
\text{,}$
%\end{equation*}
thus
%\begin{equation*}
$m=\tan(\alpha_1) > \tan(
\pi-2\arctan (1/{\sqrt{2}})-\arctan (7\sqrt{2}/8))=
\frac{23}{20}\sqrt{2}
\text{.}$
%\end{equation*}
If in \eqref{eq:y1} we do $m=\frac{23}{20}\sqrt{2}$ and $y_2 = y_2^*$ we then obtain the inferior limit for $y_1$
\begin{equation}
\label{eq:y1min}
y_1^*=
\frac{23}{10}\sqrt{2}
-\frac{1}{90}\sqrt{
444498-33120\sqrt{2}\sqrt{-51+6\sqrt{79}}
-38400\sqrt{79}
}
\simeq 1.274 \text{.}
\end{equation}
Summarizing,
%\begin{equation*}
$(y_1^*,y_2^*,y_3^*,y_4^*)\simeq
(1.274, 1.356, 0.670, 0)
\text{.}$
%\end{equation*}

With the help of illustration (b) of figure~\ref{fig:parabOpt3colB} we will, finally, analyze the entry angle $\varphi$ of the particle. With the system of equations ($\frac{1}{2}y_1=\frac{\mathrm{d} x}{\mathrm{d} y}\left.\right|_{y=y_1}$)
\begin{equation*}
\left\{
\begin{array}{l}
2\theta_1+\varphi +\alpha_1 = \pi\\
\arctan \left(\frac{1}{2}y_1\right)+\varphi +\theta_1 = \frac{\pi}{2}
\end{array}
\right.
\end{equation*}
and with equalities \eqref{eq:alpha1} and \eqref{eq:alpha2b} we obtain, successively,
\vspace{-4mm}
\begin{eqnarray}
\varphi &=&\alpha_1-2\arctan ({y_1}/{2})
\text{,}\nonumber\\
\label{eq:phiP2}
\varphi
&=&
\pi-2\arctan ({y_1}/{2})
-2\arctan ({y_2}/{2})-\alpha_2
\text{,}\\
\varphi
&=&
\pi-2\arctan ({y_1}/{2})
-2\arctan ({y_2}/{2})
-2\arctan ({y_3}/{2})+\arctan ({(y_3+y_4)}/{4})
\text{.}
\nonumber
\end{eqnarray}
Taking \eqref{eq:phiP2}, from \eqref{eq:alpha2Maj} we deduce that
%\begin{equation*}
$\varphi < \pi-2\arctan \left({y_1}/{2}\right)
-2\arctan \left({y_2}/{2}\right)
-\arctan ({(\sqrt{2}+y_3)}/{4})
\text{.}$
%\end{equation*}
In agreement with the definitions \eqref{eq:minorantes} and with the values found in \eqref{eq:y2min}, \eqref{eq:y3min} and \eqref{eq:y1min}, we can conclude that
\begin{equation*}
\varphi < \pi-2\arctan ({y_1^*}/{2})
-2\arctan ({y_2^*}/{2})
-\arctan ({(\sqrt{2}+y_3^*)}/{4})
\simeq 19.18^\circ
\text{,}
\end{equation*}
or that is
\begin{equation*}
\varphi < \varphi_0 \simeq 19.47^\circ
\text{.}
\end{equation*}
With this we can finally conclude that it is impossible to have a fourth reflection, since for this to happen the particle would have to have entered in the cavity with an angle $\varphi < \varphi_0$, as we have just finished showing --- something which would contradict our initial imposition, $\varphi > \varphi_0$. As the cavity presents symmetry in relation to its central vertical axis, the conclusion to which we have arrived is equally valid for $\varphi < -\varphi_0$, thus being proven that which we intended (theorem~\ref{teor:3ref}):
\begin{quotation}{\sl
For $\left|\varphi\right|>\varphi_0$, there always occur three reflections, alternatively on the left and right facets of the Double Parabola cavity.
}\end{quotation}

\section{Proof of theorem~\ref{teor:min3ref}}
%\section{Minimum number of reflections}
\label{cha:min3col}

For $\left|\varphi\right|>\varphi_0$ the statement of theorem~\ref{teor:min3ref} is already proved. It remains to consider the case  $-\varphi_0<\varphi<\varphi_0$, but given the symmetry of the cavity we will only need to study the interval $0<\varphi<\varphi_0$.

We will base our proof on some of the deductions which we made in appendix~\ref{cha:condSuf3col}, the illustrations of figure~\ref{fig:parabOpt3col} being especially useful to us. We will also assume to be truthful the following premise: ``If the second reflection happens on the same facet of the cavity where the first reflection occurred, there will necessarily be a third reflection''. We exempt ourselves from proving this principle because it appears evident to us.

For $0<\varphi<\varphi_0$ the first reflection can just as well occur on the left-hand facet as on the right-hand side. We will analyze each one of the cases separately.

%\vspace{0.3cm}
\noindent
\emph{1st reflection on the left-hand side}

Being $0<\varphi<\varphi_0$ we can have the first two reflections on the left-hand facet, it being in this case guaranteed, as we assume above, that $3$ or more reflections will exist. If on the other hand, the second reflection is on the right side, an initial part of the trajectory can always be represented by the first three illustrations of figure~\ref{fig:parabOpt3col} (assuming $0<\varphi<\varphi_0$), which guarantee, also in this case, the existence of a third reflection $B_3$. In order to prove what we have just finished saying, it will be enough to prove the ascendant nature of the sub-trajectory $\overrightarrow{B_1B_2}$.

We establish on the parabola of the left side (illustration~(a) of figure~\ref{fig:parabOpt3col}) the first point of reflection $B_1$. For whatever $B_1$ is it is always possible for us to trace an initial sub-trajectory $\overrightarrow{B_0B_1}$ with its origin in an entry angle $\varphi>\varphi_0$. As in appendix~\ref{cha:condSuf3col} (page~\pageref{pg:subtrajB1B2}) we showed the sub-trajectory  $\overrightarrow{B_1B_2}$ which would follow it to be ascendant, the same will necessarily come about for whatever $0<\varphi<\varphi_0$ may be, given that in this case $\overrightarrow{B_0B_1}$ will represent a more accentuated negative slope. Since in appendix~\ref{cha:condSuf3col} (page~\pageref{pg:subtrajB2B3}) we characterized $\overrightarrow{B_2B_3}$ only with basis in the ascendant nature of the sub-trajectory preceding $\overrightarrow{B_1B_2}$, the conclusions to which we arrive for $\overrightarrow{B_2B_3}$ are equally valid for $0<\varphi<\varphi_0$.

%\vspace{0.3cm}
\noindent
\emph{1st reflection on the right side}

\begin{figure}[!ht]
\begin{center}
\includegraphics*[width=0.255\columnwidth]{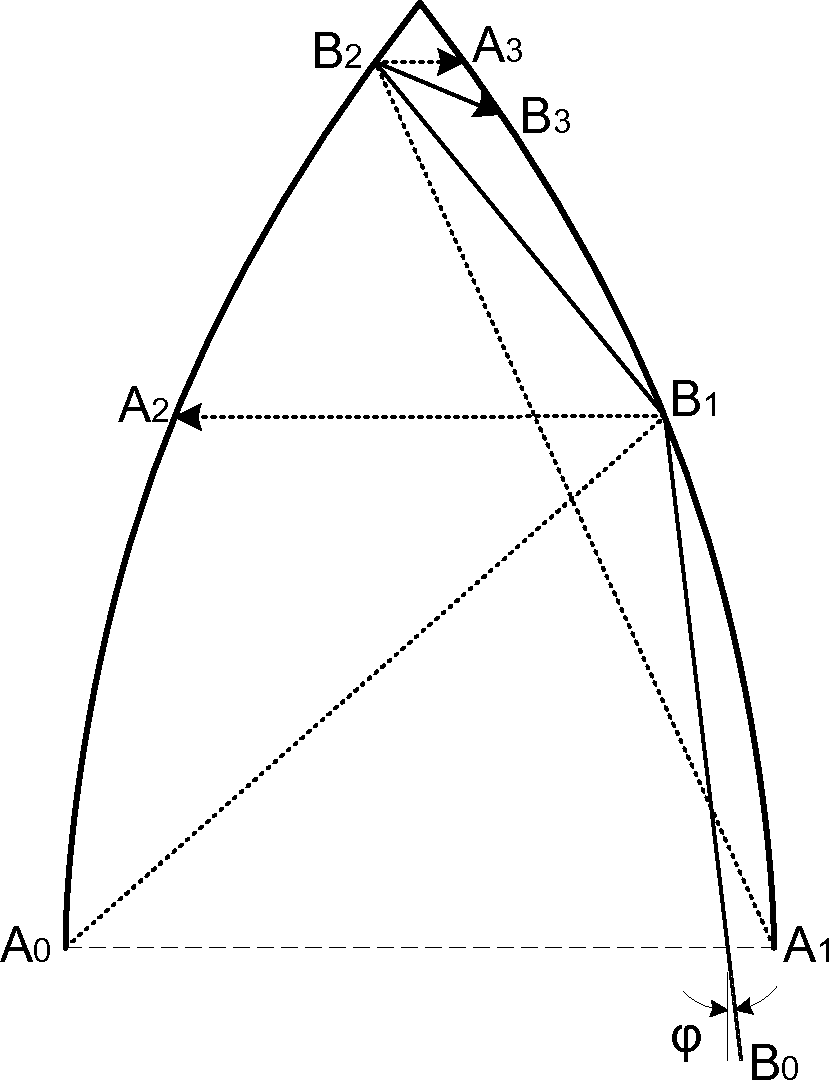}
\caption[Illustrative scheme to the study of the trajectory with entry angle $0<\varphi<\varphi_0$.]{Illustrative scheme to the study of the trajectory of particles with entry angle $0<\varphi<\varphi_0$.}
\label{fig:parabOptMin3col}
\end{center}
\end{figure}

Also in this case we can have the first two reflections on the right-hand facet, it being guaranteed that $3$ or more reflections will exist. If this does not occur, we will necessarily have a trajectory with the aspect of the trajectory $B_0B_1B_2B_3$ illustrated in the scheme of figure~\ref{fig:parabOptMin3col}, where as well there are represented two auxiliary trajectories (the dotted lines), $A_0B_1A_2$ and $A_1B_2A_3$, which, on passing through the foci of the parabolas, present the sub-trajectory posterior to the reflection horizontal.
%%%%%%%%% Figura %%%%%%%%%%%%%%%%%%%%%%%%%%%%%%%%%%%%%%%%%%%%%%%%%%%%%%%%%%%%%%
%%%%%%%%% Figura %%%%%%%%%%%%%%%%%%%%%%%%%%%%%%%%%%%%%%%%%%%%%%%%%%%%%%%%%%%%%%
%%%%%%%%% Figura %%%%%%%%%%%%%%%%%%%%%%%%%%%%%%%%%%%%%%%%%%%%%%%%%%%%%%%%%%%%%%
%%%%%%%%% Figura %%%%%%%%%%%%%%%%%%%%%%%%%%%%%%%%%%%%%%%%%%%%%%%%%%%%%%%%%%%%%%
Having as a basis the laws of reflection, we can succinctly deduce the following: as the angle $\widehat{A_0B_1A_2}$ must be interior to the angle $\widehat{B_0B_1B_2}$, we conclude that $\overrightarrow{B_1B_2}$ is of a ascendant nature;
as $\widehat{B_1B_2B_3}$ is necessarily an interior angle to $\widehat{A_1B_2A_3}$, we conclude that $B_3$ must be situated between $A_1$ and $A_3$, which guarantees the existence of a third reflection. Thus is the proof of the theorem~\ref{teor:min3ref} concluded.
%%%%%%%%%%%%%%%%%%%%%%%%%%%%%%%%%%%%%%%%%%%%%%%%%%%%%%%%%%%%%%%%%%%%%%%%%%%%%%%
%%%%%%%%%%%%%%%%%%%%%%%%%%%%%%%%%%%%%%%%%%%%%%%%%%%%%%%%%%%%%%%%%%%%%%%%%%%%%%%
%%%%%%%%%%%%%%%%%%%%%%%%%%%%%%%%%%%%%%%%%%%%%%%%%%%%%%%%%%%%%%%%%%%%%%%%%%%%%%%

\section*{Acknowledgements}
This work was supported by the
\emph{Centre for Research on Optimization and Control} (CEOC)
from the \emph{Portuguese Foundation for Science and Technology} (FCT),
cofinanced by the \emph{European Community Fund} (ECF)
\textsf{FEDER/POCI 2010}; and by the FCT research project
\textsf{PTDC/MAT/72840/2006}.

%\bibliographystyle{unsrt} %paulo
%\bibliography{resist}

\end{document}